\documentclass[journal]{vgtc}                     % final (journal style)
\onlineid{1100}

\vgtccategory{Research}

\vgtcpapertype{algorithm/technique}

\title{GSCache: Real-Time Radiance Caching for Volume Path Tracing\\using 3D Gaussian Splatting}

\author{%
  \authororcid{David Bauer}{0000-0002-1327-3054},
  \authororcid{Qi Wu}{0000-0003-0342-9366},
  \authororcid{Hamid Gadirov}{0000-0001-6578-4342}, and 
  \authororcid{Kwan-Liu Ma}{0000-0001-8086-0366}
}

\authorfooter{
  \item
  	David Bauer and Kwan-Liu Ma are with the University of California at Davis.
  	E-mails: \{davbauer, klma\}@ucdavis.edu
    \item Qi Wu is with NVIDIA Research
        E-mail: qiwu@nvidia.com
    \item Hamid Gadirov is with the University of Groningen
        E-mail: h.gadirov@rug.nl
}

\abstract{%
Real-time path tracing is rapidly becoming the standard for rendering in entertainment and professional applications. In scientific visualization, volume rendering plays a crucial role in helping researchers analyze and interpret complex 3D data. Recently, photorealistic rendering techniques have gained popularity in scientific visualization, yet they face significant challenges. One of the most prominent issues is slow rendering performance and high pixel variance caused by Monte Carlo integration. In this work, we introduce a novel radiance caching approach for path-traced volume rendering. Our method leverages advances in volumetric scene representation and adapts 3D Gaussian splatting to function as a multi-level, path-space radiance cache. This cache is designed to be trainable on the fly, dynamically adapting to changes in scene parameters such as lighting configurations and transfer functions. By incorporating our cache, we achieve less noisy, higher-quality images without increasing rendering costs. To evaluate our approach, we compare it against a baseline path tracer that supports uniform sampling and next-event estimation and the state-of-the-art for neural radiance caching. Through both quantitative and qualitative analyses, we demonstrate that our path-space radiance cache is a robust solution that is easy to integrate and significantly enhances the rendering quality of volumetric visualization applications while maintaining comparable computational efficiency.
}

\keywords{Radiance caching, path tracing, volume rendering, gaussian splatting}

\teaser{
  \centering
  \includegraphics[width=\textwidth]{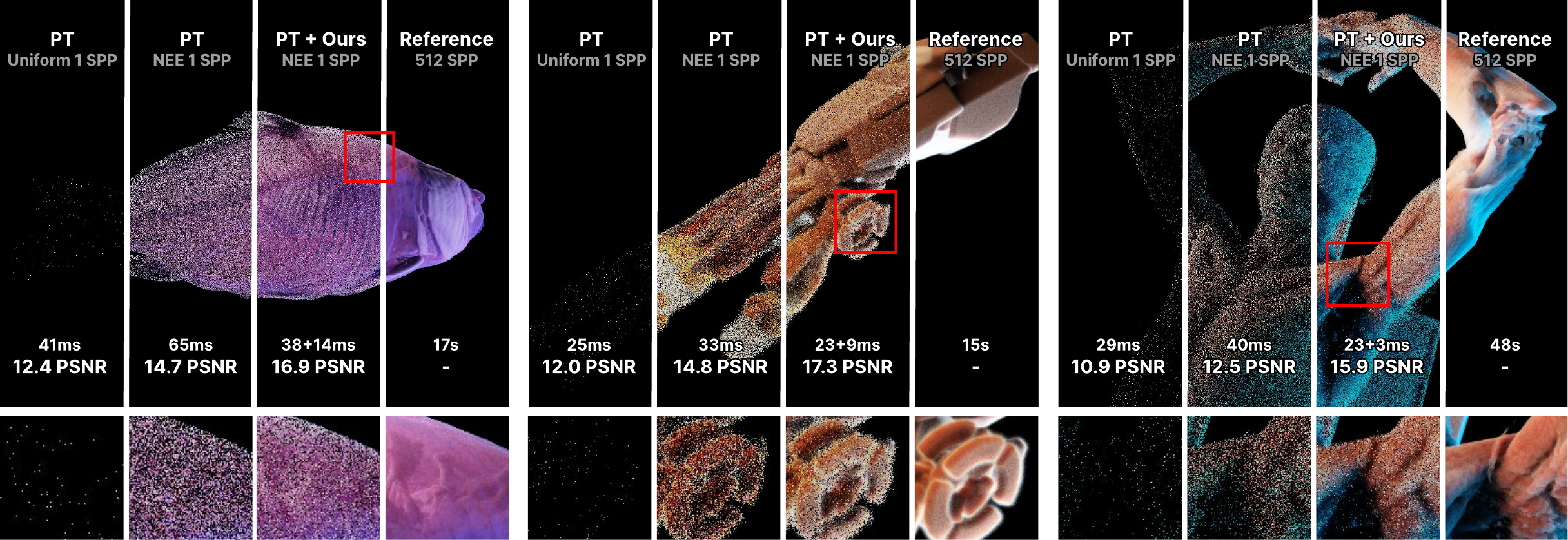}
  \caption{%
     Our path-space cache improves image quality at low sample counts at comparable compute cost. We compare against a path tracer (PT) using uniform sampling and a version implementing next-even-estimation (NEE). Cache rendering time is constant, yielding increasing returns as sample counts increase. The method is non-invasive and easy to integrate into existing rendering applications.
  }
  \label{fig:teaser}
}

\graphicspath{{figs/}{figures/}{pictures/}{images/}{./}} % where to search for the images

\usepackage{tabu}                      % only used for the table example
\usepackage{booktabs}                  % only used for the table example
\usepackage{lipsum}                    % used to generate placeholder text
\usepackage{mwe}                       % used to generate placeholder figures

\usepackage{mathptmx}                  % use matching math font

\usepackage{lmodern} % Allows arbitrary font size and removes OT1/cmr/bx/n warning
\usepackage{multirow}
\usepackage{algorithm}
\usepackage{algpseudocode}
\usepackage{amsmath}
\usepackage{tabularx}
\usepackage{caption}
\usepackage{subcaption}
\usepackage{graphicx}
\usepackage[svgnames]{xcolor}
\usepackage{soul}

\newif\iftodovisible
\todovisibletrue
\definecolor{todotext}{RGB}{0,0,0}
\definecolor{todobg}{RGB}{255, 219, 168}

\newif\ifcheckvisible
\checkvisiblefalse
\definecolor{checktext}{RGB}{0,0,0}
\definecolor{checkbg}{RGB}{255, 50, 50}

\newif\ifdiffvisible
\diffvisiblefalse
\definecolor{diffdel}{RGB}{255,0,0}
\definecolor{diffadd}{RGB}{0,128,0}
\newcommand\revdel[1]{\ifdiffvisible{\color{diffdel}\st{#1}}\fi}
\newcommand\revadd[1]{\ifdiffvisible{\color{diffadd}#1}\else#1\fi}

\begin{document}
\maketitle

%% Sections 
\section{Introduction}
Photorealistic rendering of scientific volume datasets in real time has become a feasible alternative to more traditional rendering techniques in recent years. The advances in rendering research and graphics hardware have enabled real-time path tracing for many visual effects. These applications rely on low sample counts, smart sampling strategies, and advanced post-processing techniques to provide acceptable output quality. 

Despite this, high-quality real-time rendering remains challenging in domains like volume rendering, which can be prone to excess levels of Monte Carlo noise as a result of high sampling variance. Noise mainly stems from variance in the sampling distribution and the need to sample particle interactions with the volume---a process not required in surface-only path tracers. In extension, variance is influenced by the quality of the samples we take. Past research has focused on different ways to address these issues by improving the sampling scheme to create more relevant samples, making the few samples we take count more~\cite{restir, volumerestir, gris, restirgi}. \revdel{On the other hand,}\revadd{Orthogonal approaches like} radiance caching \revdel{methods} can be used to store previously seen radiance samples in a cache data structure, which can then be used to augment subsequent samples~\cite{ward88,krivanek05radiance,muellernrc,Hadadan2021NeRad}. Many times, these methods are highly specialized solutions that require much additional work to be integrated into existing renderers. Despite these advances, scientific volume rendering has received little attention in these domains, making it an interesting subject for further improvement.

In this work, we introduce a real-time path-space cache for volume rendering that stores attenuated radiance at different path lengths and can quickly adapt to changes in the scene, such as lighting, transfer function, or slicing operations. The cache is represented as a multi-level collection of 3D Gaussians. Each level caches the attenuated path-space radiance for paths of a specific length. Cache sizes are sub-sampled from higher to lower levels, similar to the construction of MIP levels in texture processing. This is done since paths of length $n$ generally contribute more to the final image than paths of length $m$, where $m > n$. This property allows us to allocate more computational resources and representational power to paths that have a higher image contribution. The cache is optimized in real time with noisy samples from the renderer. The only information needed from the rendering application is the attenuated radiance value and the path length that produced it.

We evaluate our approach by comparing it to a baseline volume path tracer. In addition to image quality and rendering speed analysis, we conduct an ablation study to justify our design. We test our approach on a scientific volume renderer that implements path tracing with geometry-based lighting and compare results from uniform sampling and using next-event estimation (NEE). Our results show that Gaussian radiance caches are an effective, fast, and easy-to-integrate alternative to world- and image-space-based radiance caches. Our contributions can be summarized as follows.

\begin{itemize}
    \item We introduce a novel radiance cache optimized for volume rendering that caches path-space radiance using multiple levels of Gaussian splats.
    \item The cache works in real time on complex datasets and in a wide variety of use cases and adapts quickly to changes in the transfer function and lighting parameters, which improves the overall image quality and rendering times.
    \item Optimizing the cache is possible not only with clean samples but also with noisy data, as is commonly found in Monte-Carlo-based renderers.
    \item The path-space nature of the cache and its non-invasive design make it easy to use and integrate into existing rendering solutions.
\end{itemize}

With these contributions, we hope to support the design and quality of scientific renderers and encourage the use of radiance caching in scientific applications and beyond.
\section{Related Work}
This work is related to methods in scientific visualization, volume rendering, radiance caching, and machine learning. In the following paragraphs, we discuss relevant works from these fields. 

\subsection{Radiance Caching}
Radiance caching aims to trade variance for bias to improve the image quality of path tracing applications. In the context of path tracing, variance is the result of sampling decisions along a path, while bias is generally the result of making assumptions about radiance distributions in a scene that are not analytically tractable. \revdel{Traditionally,}\revadd{In most cases} (ir-)radiance is cached \revadd{by storing radiance probes} in world-accessible spatial data structures that can be sampled during rendering~\cite{krivanek05radiance,jarosz08radiancemedia,stadlbauer2025}. \revdel{Newer}\revadd{Recently, some} approaches have made use of deep learning to store and access radiance information~\cite{muellernrc, bauer24photon, Hadadan2021NeRad}. Early approaches usually focus on caching indirect illumination after the first bounce. This means that all paths that terminate into the cache do so after a single bounce. A downside of such approaches is that using the cache so soon can introduce considerable error from bias. This is why other approaches try to mask the bias by allowing longer paths~\cite{keller14pathspacefiltering, muellernrc}. \revadd{There has also been work specifically targeting volume rendering. Khlebnikov et al.~\cite{khlebnikov2014parallelirradiance} introduce a parallel GPU-based irradiance caching approach for interactive volume rendering. Jakob et al.~\cite{jakob2011progressive} use hierarchical Gaussian mixture models to derive scene radiance from photon traces and \v{S}majdek et al.~\cite{smajdek2024combinedvolsurfcache} introduce radiance caching to address mixed-modality scenarios where isosurface and volume rendering are combined.}

One crucial consideration to make when allowing for different path lengths is to decide when to terminate a path and instead to read stored radiance from the cache and update its entries. Throughout the literature, several heuristics have been developed to address this issue. For example, Müller et al.~\cite{muellernrc} terminate paths based on their spatial spread using a heuristic proposed by Baekert et al.~\cite{bekaert2003custom}. Kandlbinder et al.~\cite{Kandlbinder2024}\revdel{, on the other hand,} develop a variance-based heuristic that offers better control of the bias error bound when sampling the cache.

\subsection{Scientific Volume Rendering}
Volume rendering is vital to the field of scientific visualization. Broadly speaking, applications of scientific rendering can be categorized into two groups---visualization for analysis and visualization for illustration. The former group relies on rendering techniques that offer interactive framerates and user interactions, such as cutting planes and transfer function editors. We commonly see direct volume rendering (DVR) as the rendering method of choice in this group. Illustrative visualization, on the other hand, prioritizes image quality over interactivity and customization. It is closer, in purpose, to more traditional offline rendering scenarios. In this case, path tracing with physically-based illumination models is preferable.
Both types of rendering and the underlying optical models for volume visualization have been aptly summarized by Max et al.~\cite{max1995optical}. Despite promises of realism, calculating physically-based volumetric illumination via the radiative transfer equation (RTE) is expensive. For a long time, research has focused on finding appropriate approximations to enable real-time rendering of scientific datasets. J\"onsson et al.~\cite{joensson2014interactivedvr} provide an excellent summary of this category of techniques. In the following, we highlight some notable works in this area.
Gradient-based methods were one of the first ways to render volume datasets. Levoy et al.~\cite{levoy88surfacefromvolume} use volume gradients to apply surface shading models like the Blinn-Phong reflection model. Local shading is a simple step up from gradient-based rendering. Ambient occlusion techniques~\cite{ruiz2008obscurance, schott2009directional, vsolteszova2010multidirectional} have been popular in the domain as they are relatively easy to implement and grant viewers a greater sense of depth, which makes it easier to discern details in complex or abstract volume datasets. Next, slice-based methods like Kniss et al.'s half-angle slicing~\cite{kniss2002interactive,kniss2003model} provide a more sophisticated model and were one of the first to simulate multi-scatter illumination.
Lastly, physically-based particle tracking~\cite{woodcock1965} and path tracing allow for high levels of realism with full global illumination. Here, Zhang et al.~\cite{zhang2013prephoton} use pre-computed photon maps while others~\cite{dappa2016cinematic,kroes2012exposure} work on ray tracing-based approaches. J\"onsson et al.~\cite{jonsson2012historygrams} expand on the idea of photon mapping for scientific visualization and introduce an approach that selectively recomputes photon traces based on transfer function changes. Lastly, Bauer et al.~\cite{bauer24photon} transform photon maps into implicit neural representations, which improves rendering speeds, enabling interactive visualization. 
% More recent advances in GPU hardware, machine learning, and rendering research have tremendously improved the cost and practicality of real-time path tracing application. This trend has seen path tracing emerge as the general tool of choice for cinematic~\cite{dappa2016cinematic} scientific volume rendering.
% Despite these advances, volume path tracing for interactive applications remains challenging. In contrast to DVR, path tracing generally produces noisy images due to sample variance from Monte Carlo integrations. For scientific visualization, there have been efforts to use precomputed radiance volumes~\cite{zhang2013prephoton,bauer24photon} to alleviate some of this noise and provide high-fidelity images at interactive framerates.
Our work is related to these techniques in that we aim to provide higher performance and lower variance for scientific path tracing applications.

\subsection{Machine Learning for Scientific Visualization}
Scientific visualization research has been comparatively slow to adapt machine learning and deep learning into its repertoire. Nevertheless, recent years have witnessed several interesting applications of machine learning techniques to address pervasive problems in scientific visualization and rendering. 

Finding good transfer functions and viewing parameters has been a longstanding challenge in scientific visualization. Weiss et al.~\cite{weiss22diffvol} introduce a differentiable direct volume rendering algorithm that addresses this issue. Using their method, parameters like the input transfer function, volume density, or camera configuration can be optimized via gradient descent. Later, Pan et al.~\cite{pan2023diffdesigngalleries} expand on this idea and introduce an interactive tool to explore the latent space of different transfer function designs. Users can modify designs and optimize for the corresponding transfer function using Weiss et al.'s~\cite{weiss22diffvol} approach.

Real-time rendering is particularly challenging in scientific visualization as datasets tend to be large and complex. Developing rendering applications for such data requires careful design to operate within the memory and performance limits. This becomes especially pressing when advanced illumination models like ambient occlusion (AO) or multi-scatter path tracing are used. Engel et al.~\cite{engel2020deep} introduce a neural representation for real-time volumetric ambient occlusion rendering, improving baseline costs of calculating AO. Bauer et al.~\cite{bauer24photon} extend this thought to full global illumination. Their method introduces a phase-function varying neural representation of volumetric lighting that can be used to replace traditional path sampling beyond the first interaction, reducing noise and improving rendering speeds. Aside from caching approaches, there have been efforts~\cite{Weiss2022,bauer2023fovolnet} to reduce rendering costs by adaptively sampling and reconstructing data in the volume rendering pipeline.

Lastly, storage poses additional challenges over traditional rendering applications. Volume datasets tend to be large, with multiple parameter grids and even multiple volumes composed as a time series. Neural networks have proved to be a powerful tool to implicitly represent and compress such volumes~\cite{han23coordnet, weiss22vnr, wu24vnr}. Weiss et al.~\cite{weiss22vnr} and Wu et al.~\cite{wu24vnr} develop implicit neural representations (INR) for volume compression to address this issue. Zavorotny et al.~\cite{zavorotny2025cinr} address INR rendering bottlenecks while, more recently, Wu et al.~\cite{wu23hyperinr} and Gadirov~\cite{gadirov2025hyperflint} have extended neural visualization methods using hypernetworks for parameter-space exploration.
\begin{figure*}[!htb]
    \centering
    \includegraphics[width=0.85\textwidth]{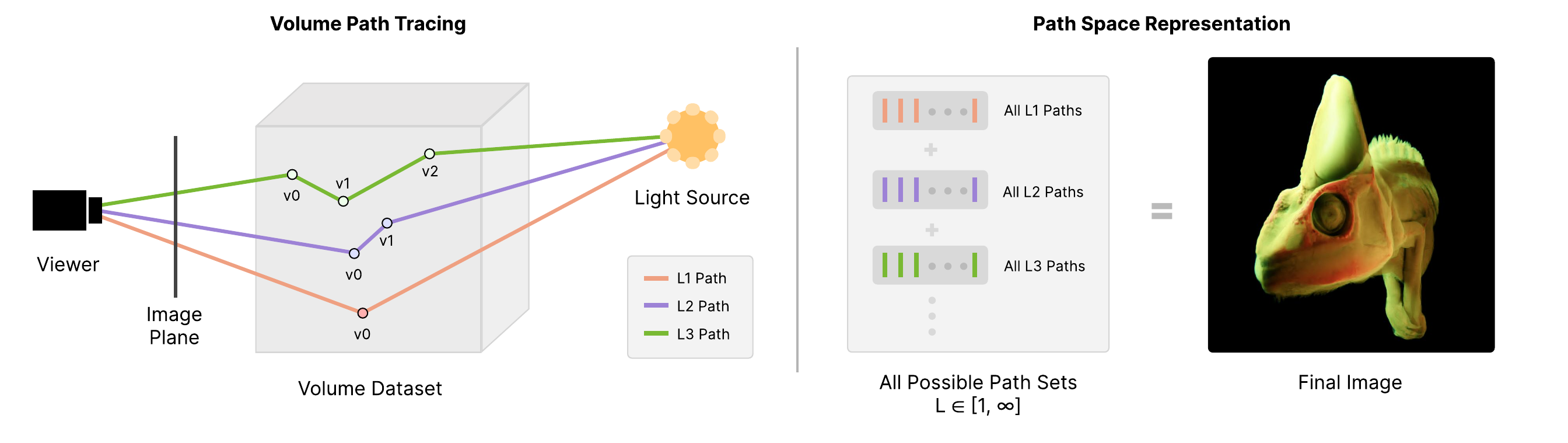}
    \vspace{-0.2in}
    \caption{The path tracing integral can be characterized as the collection of all possible paths of all possible lengths. When combined, these paths form the final image. Left: Example of a volume path tracer. Primary rays are produced by the viewer and interact with the volume at path vertices $v_i$. The number of interactions with the volume before a path terminates determines the path length. Right: Paths of the same length can be grouped into path-space sets.}
    \label{fig:path_space_rendering}
\end{figure*}

\section{Method}
Our method broadly consists of two parts. The first is a path-space radiance representation attained by training our Gaussian cache. The second component is the integration of our cache into a volume path tracer for real-time rendering and cache optimization. This section describes our cache design in detail.

% \todo{
%     Here we introduce the concept of using Gaussian splats as path-space radiance caches. We describe the general mechanism and how it extends to multiple levels to capture path space.
% }

\subsection{A Path Space Radiance Cache}
% \todo{discuss concept of path space radiance and how we can cache it using GSRC}

Traditionally, the volumetric radiance transfer can be characterized by the integral of transmittance-attenuated radiance accumulated along a sampling ray (see Equation~\ref{eq:volrad_transfer}).

\begin{equation}
    L(x,\omega) = \int_0^\infty{T_r(x^\prime \to x) L_s(x^\prime,-\omega)dt}
    \label{eq:volrad_transfer}
\end{equation}

\noindent where $T_r$ is the transmittance between two points $x$ and $x^\prime$ and $L_s$ is the source term describing the amount of radiance going through point $x$ in direction $-\omega$. This radiance integral can be reformulated as an integral in path space. In this formulation of radiance, we consider sets of possible paths of length $n$. The final color of a pixel can be found by adding up the contribution of all path sets $n \in [0, \infty]$ (see Figure~\ref{fig:path_space_rendering}).
% \begin{equation}
%     I = \sum_{i=0}^{\infty}{\frac{1}{N}\sum_{j=0}^{\infty}{I_j}}
%     \label{eq:path_space_sum}
% \end{equation}
% \todo{Path space sum equation is not complete}
For a detailed examination of the path space radiance formulation, please refer to Chapter 8 in Veach~\cite{veach1998robust}. Our cache design is based on this same notion of path space radiance. We represent the total radiance of all paths of length $n$ in a single cache level. For each $i \leq n$ we define a separate cache level to hold the radiance of that subspace of path space radiance (see Figure~\ref{fig:cache_sampling}).

Each cache level is represented as a point cloud of three-dimensional Gaussians. These Gaussians can be used to reconstruct the path space subset radiance for each level by rasterizing them into images using the recently introduced Gaussian splatting rasterization technique~\cite{kerbl3Dgaussians}. One notable feature of 3D Gaussian splatting is its fast differentiable rasterizer, which allows for splatting at high frame rates. We make use of this property to enable fast access to radiance at each cache level. Before every path tracing pass, a new set of cache images is rasterized and passed to the path tracing algorithm. After one pass in the path tracer, the cache is optimized with fresh path samples. In the following sections, we describe how the cache is initialized, how paths are terminated into the cache, and how it is optimized in real time during rendering.

\subsection{Cache Initialization}
\label{sec:cache_init}
We initialize the radiance cache by doing a single sampling pass over the volume data. The goal of this step is to generate a point cloud that will be used to initialize each cache level with a set of Gaussians that roughly correspond to the structure of the dataset. To this end, we randomly generate rays that intersect the volume. The rays are traced through the dataset using delta tracking~\cite{woodcock1965}. If there is an interaction, we record the position and albedo of the interaction and terminate the process. In cases where the tracking algorithm does not produce a valid interaction (i.e., when the ray leaves the volume), we repeat the process with a newly generated random ray until we find a valid point sample.
For our purposes, we generate an initial set of $N$ points to initialize the cache. In our tests, we found $N = 300$k to be a suitable initialization size. Upon initialization, the point cloud is replicated and logarithmically sub-sampled for each level, allotting more points to earlier levels in the cache. This process results in a total number of $\sum_{i=0}^{K}{\frac{N}{2^i}}$ points where $K+1$ is the number of levels chosen for the cache. Based on the point locations and colors, we initialize a set of Gaussians similar to Kerbl et al.~\cite{kerbl3Dgaussians}. We use a k-nearest neighbor query to determine the initial scales of the Gaussians. Unlike Kerbl et al.~\cite{kerbl3Dgaussians}, we use a smaller initial scale for the Gaussian covariances. Additionally, we cap any outliers with a z-score greater than $2$ to at most two standard deviations above the mean to exclude unreasonably large Gaussians in our starting condition.

\begin{equation}
    s_i = \frac{(\mu_N + 2*\sigma_N) \land \frac{1}{3}\sum_{j=0}^{2}{d_{ij}}}{2}
\end{equation}

\noindent where $s_i$ is the $i$-th Gaussian's isotropic scale, $d_{ij}$ is the distance to $j$-th closest neighbor of the $i$-th point, and $\mu_N$ and $\sigma_N$ are the mean and standard deviation over all $N$ points' 3-closest neighbor distances. Effectively, this process eliminates large outliers and scales the overall set of Gaussians to $50\%$. Leaving enough space between Gaussians helped stabilize cache optimization on noisy input data.

\subsection{Path Termination Heuristics}
For caching to be efficient, it is common to introduce a path termination heuristic to determine if a path will contribute a meaningful amount of radiance. Several heuristics have been used in surface path tracing for radiance caching~\cite{Kandlbinder2024,bekaert2003custom}. Notably, Kandlbinder et al.~\cite{Kandlbinder2024} introduced effective heuristics to terminate paths that are unlikely to contribute significant radiance to the integral. M\"uller et al.~\cite{muellernrc} used methods by Bekaert et al.~\cite{bekaert2003custom} for neural radiance caching. Unlike much of prior work~\cite{krivanek05radiance,jarosz08radiancemedia,muellernrc,bauer24photon,Hadadan2021NeRad}, our method operates in path-space. In addition to that, our cache is specifically targeted at volume rendering applications---an area that has received little attention in prior work. These facts change the functional requirements of the heuristic. 

The overall goal of such a path termination heuristic is to determine when the benefits of tracing a full, unbiased path outweigh the benefits of using a biased cache sample instead. Traditionally, an effective measure to determine this has been to estimate the current path's contribution if it were to leave the volume after the current path vertex to receive radiance from the environment outside the volume. In the case of basic uniform path tracing, this estimation occurs when the path naturally terminates, which happens when the tracking algorithm does not find a valid interaction with the volume, indicating that the path has left the volume bounds. This approach is easy to implement but suffers from the fact that paths have to be traced to their natural conclusion, foregoing the potential savings of terminating a low-contribution ray early. To improve upon the na\"ive approach, we leverage the commonly implemented technique of next-event-estimation (NEE). This technique estimates direct light contribution at every path vertex. The light sample is then combined with the continued path contribution using multiple importance sampling. Having direct light samples at every path vertex allows us to continuously estimate a path's contribution as we trace it through the volume. In our implementation, we distinguish two cases---natural and early path termination---and we propose two simple cache sampling strategies for both cases.

\subsubsection{Natural Path Termination}
The base case in most volume path tracers is reached when there is no new valid interaction with the volume, and the path naturally leaves the volume.
Since path termination is not deliberately sampled but rather an unforeseeable outcome of phase function sampling, terminating directions are chosen according to the volume's scattering behavior. The majority of such terminal path segments typically does not intersect with any of the light sources in the scene, leading to zero radiance contribution along the ray, effectively wasting the whole path sample. In scenes with infinite area light sources, samples might contribute a small amount of radiance.
We can make use of this fact to sample our cache. If the path terminates with a non-zero radiance contribution, we use the unbiased sample; otherwise, we determine the cached radiance at the terminating path length and use this value instead. In this way, every path that terminates contributes a meaningful amount of radiance to the final image.

\subsubsection{Early Path Termination}
\label{sec:early_path_termination}
In addition to sampling the cache for naturally terminated paths, we also consider terminating paths early into the cache. This has the benefit that we not only get a less noisy sample but also do so at a performance gain, as reading the cache is cheaper than sampling the path to its natural termination.

The heuristic uses the luminance of the current path throughput $Tr$ to generate a termination probability $p$ (see Algorithm~\ref{alg:termination}). \revadd{The Algorithm takes a list of path vertex albedos $\sigma_i$ to compute $Tr$. At the same time, we keep track of $\beta_i$ to importance sample cache hits. The details of the algorithm are covered in the subsequent paragraphs.} The heuristic is similar to the Russian Roulette termination technique that is commonly used in path tracers to cut long paths with little contribution short. Unlike Russian Roulette, our method uses a much higher initial throughput threshold to become active. In our tests, we consider any throughput values less than $0.9$. The reason for having this threshold is to guarantee that a fraction of paths will continue regardless of any cache sampling probability to ensure that we can gather paths of all lengths. Additionally, we employ a user-defined termination coefficient $C$, which modifies the baseline probability in favor or against the current throughput. The final probability is used to determine if the current path should be terminated into the cache and, if not, is used to importance sample the continuation of the path (see Section~\ref{sec:cache_sampling}).
% \todo{Should ray termination be determined by a threshold instead of using the cache directly?}
% \todo{There should be a distinction between path termination probability and then cache sampling probability. If they are somehow the same thing, then we need to make it clear that every terminated path that has not naturally left the volume will be using the cache.}
\begin{algorithm}
\caption{Early path termination heuristic} 
\label{alg:termination}
    \hspace*{\algorithmicindent}\textbf{Input} $[\sigma_1, ..., \sigma_n]$, $C$, $\beta_n$ 
    
    \hspace*{\algorithmicindent}\textbf{Output} $Tr_{out}, \beta_{n+1}$
\begin{algorithmic}
    \State \emph{$Tr_{out}$} $\gets \prod_{k = 1}^{n}\sigma_k$
    \State \emph{$Tr$} $\gets \text{clamp}(C\cdot \text{luminance}(Tr_{out}), 0, 1)$
    \If{\emph{$Tr < 0.9$}}
        \State \emph{$q \sim \mathcal{U}(0,1)$}
        \State\emph{$p$} $\gets 1 - Tr$
        \If{\emph{$q < p$}} 
        
            \Return True
        \EndIf
        \State \emph{$Tr_{out}$} $\gets \frac{Tr_{out}}{Tr+\epsilon}$
        \State \emph{$\beta_{n+1}$} $\gets \beta_n Tr$
    \EndIf  
    
    \Return False
\end{algorithmic}
\end{algorithm}

\subsection{Cache Sampling}
\label{sec:cache_sampling}
% \todo{
%     Here we talk about how we sample the cache. Currently, the best way to sample it has been to create a probability based on the sampled, attenuated radiance and the cached radiance. 
%     The probability then decides if we should pick the cache or the sample.
% }
% \todo{ 
% Talk about how we cascade the termination probability to re-attenuate the cache samples.
% }

\begin{figure}[!htb]
    \centering
    \includegraphics[width=1.0\columnwidth]{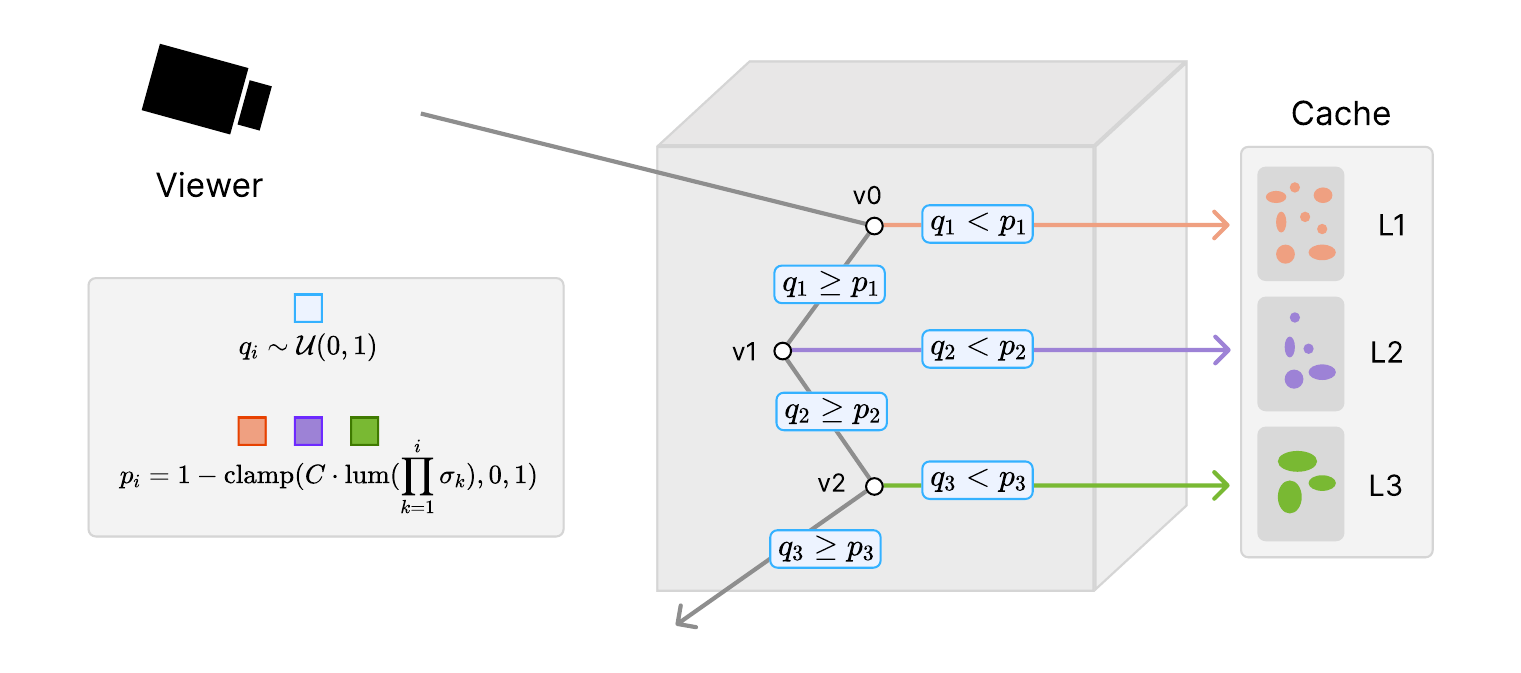}
    \caption{Illustration of our cache sampling and path termination mechanisms. Using our path termination heuristic allows us to sample the cache at different levels depending on the throughput as the product of all prior path vertex albedos $\sigma_i$ and the cache sampling coefficient $C$.}
    \label{fig:cache_sampling}
\end{figure}

% If a path is terminated at depth $n$, we want to determine if the final radiance should come from the cache or the path's contribution until vertex $n$. We sample the radiance cache based on the anticipated path contribution as represented by the current total path throughput $Tr$. 
If a path is terminated at depth $n$, we want to determine if the final radiance should come from the cache or the path's contribution until vertex $n$. To make this decision, we look to the current total path throughput $Tr$ as an indicator of anticipated path contribution (Figure~\ref{fig:cache_sampling}). We sample the radiance cache based on the anticipated path contribution to select either the value in the cache at the corresponding level or use the current unbiased radiance that we accumulated through next-event estimation. 

\subsubsection{Throughput Importance Sampling}
The considerations we make for natural path termination are sufficient to sample the cache correctly. However, when paths are terminated into the cache early, we need to make adjustments to keep the overall path sampling unbiased. First, we adjust the throughput $Tr_{out}$ for cases where the cache sampling resulted in no hit. This is equivalent to adjusting the path throughput in Russian roulette. The effective path throughput $Tr_{out}$ is importance sampled by dividing by the cache sampling probability $Tr$ (see Algorithm~\ref{alg:termination}). Aside from this, we need to consider what happens when there is a cache hit. In the following, we describe how we ensure that cache reads contribute an appropriate amount of radiance to the sample.

\subsubsection{Cascaded Cache Radiance Importance Sampling} 
\begin{figure}[!htb]
    \centering
    \begin{subfigure}[b]{.60\columnwidth}
    \includegraphics[width=\linewidth]{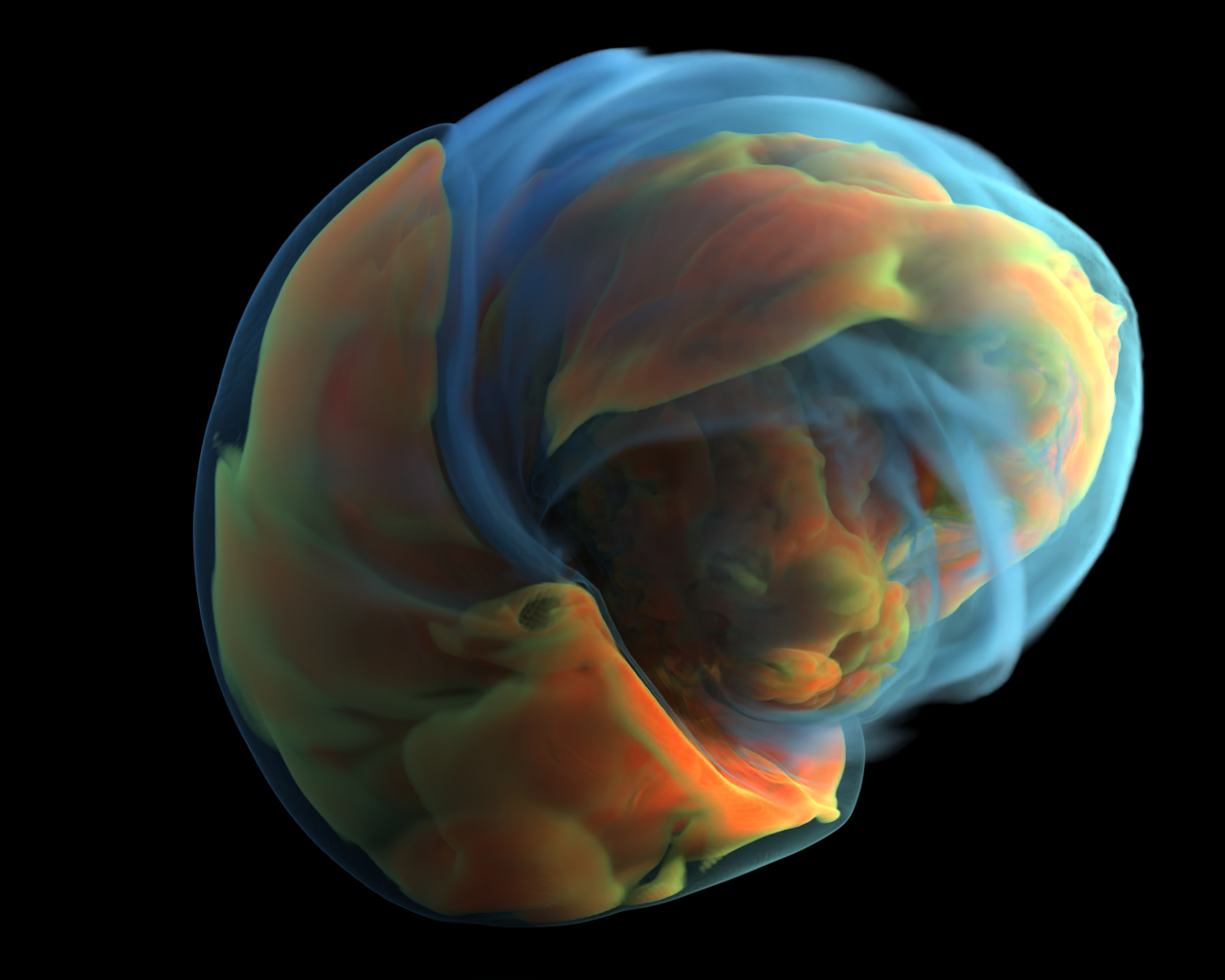}
    \end{subfigure}
    \begin{subfigure}[b]{.298\columnwidth}
        \begin{subfigure}[b]{1.0\columnwidth}
            \includegraphics[width=\linewidth]{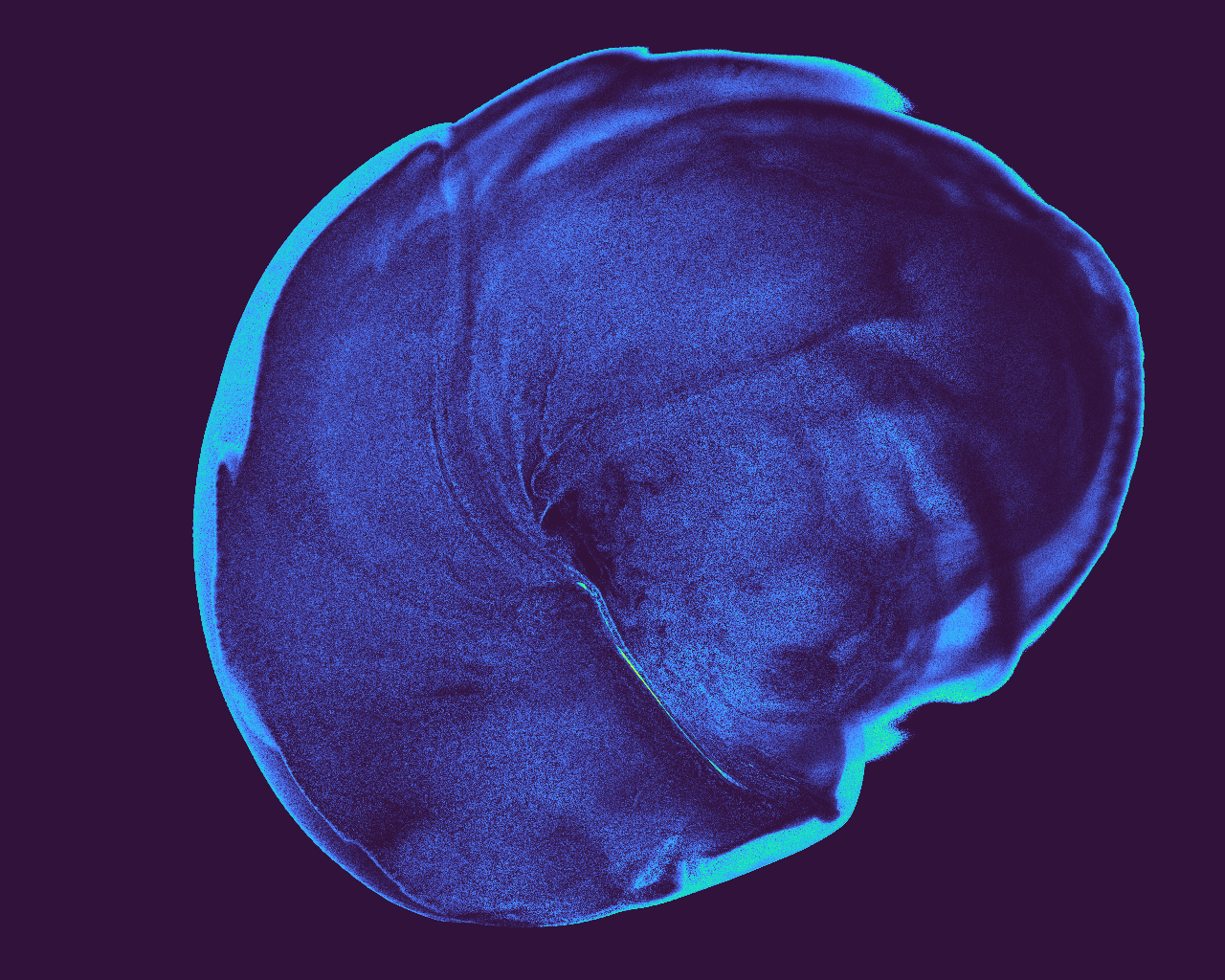}
            % \caption{rMSE error map without using $\beta$}
        \end{subfigure}
        \begin{subfigure}[b]{1.0\columnwidth}
            \includegraphics[width=\linewidth]{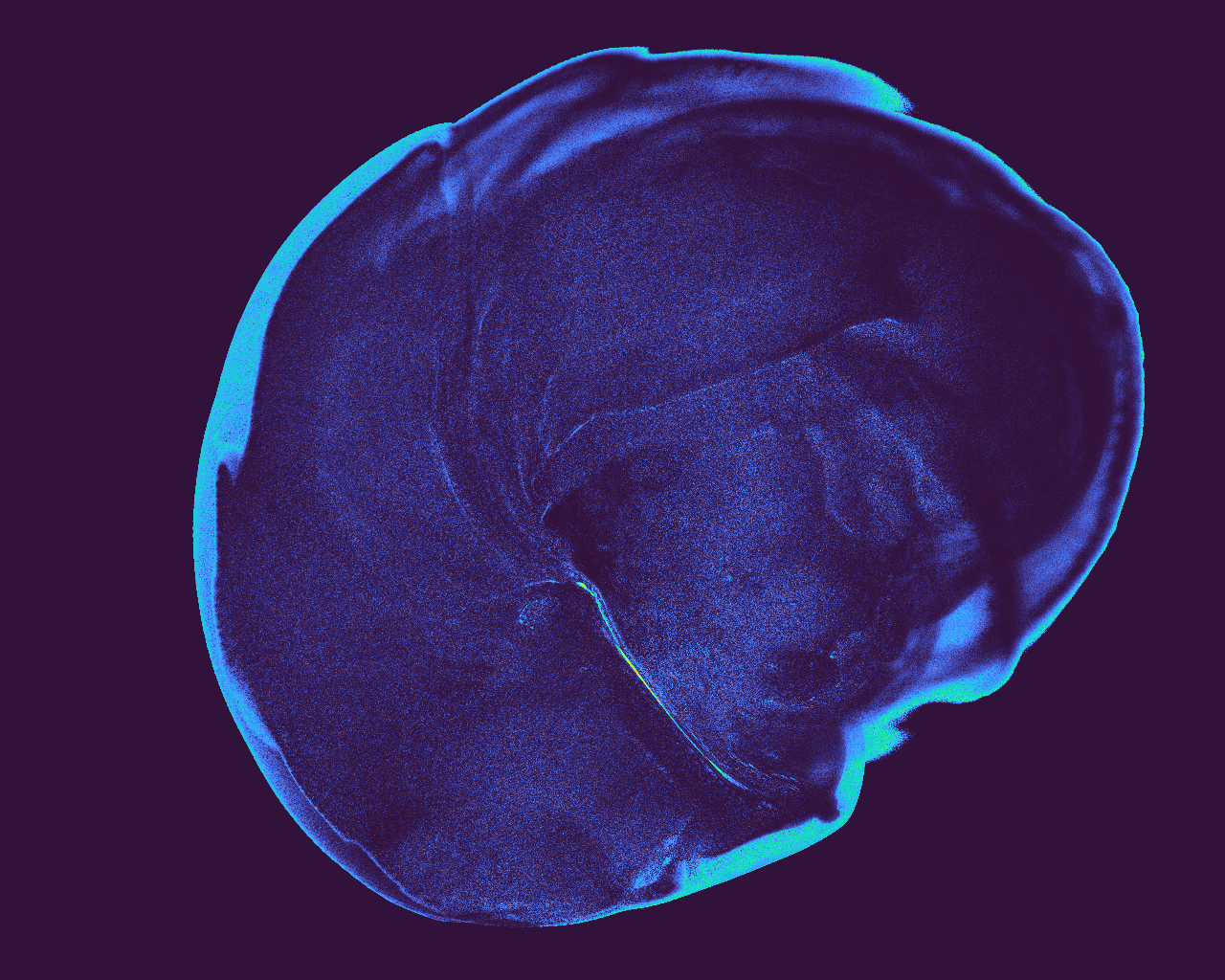}
            % \caption{rMSE error map with correct attenuation}
        \end{subfigure}
    \end{subfigure}
    \caption{Accounting for the cascaded cache sampling probability is necessary to accurately attenuate cached radiance samples. (Left) Reference image. (Top) rMSE error map of the rendering disregarding cascaded cache sampling probability $\beta$. (Bottom) rMSE error map of the rendering with correct $\beta$ attenuation. Correcting for $\beta$ clearly improves the image error in the lower image compared to the converged reference.}
    \label{fig:cascaded_probability}
\end{figure}
Adjusting path throughput $Tr_{out}$ by the probability of a cache hit is not enough to ensure correct cache sampling. Another issue arises when there is a cache hit. The cached values are pre-attenuated and directly represent the radiance at path length $n$. However, this value does not take into account the possibilities of early terminations at length $i \in [1, n-1]$ (Figure~\ref{fig:cascaded_probability}). Since paths could have terminated at any path vertex before the current one, only a fraction of paths reach length $n$. We need to account for this fact by keeping track of the product of all prior cache sampling probabilities $Tr$ and adjusting the cached radiance by this product (see Algorithm~\ref{alg:termination}). Not doing so will result in images that are generally darker than the unbiased reference.
To keep track of the product path sampling probability $\beta$, we initialize it to $1$ and multiply it by $Tr$ any time the cache is missed. When the cache is hit at depth $n$, we divide the cached radiance by $\beta_{n-1}$ to account for the fact that the cache was missed $n-1$ times. The effective attenuation applied to a path sample is as follows.

\begin{equation}
    \hat{L}_{n} = \frac{L_n \prod_{k=1}^{n}{\sigma_k}}{\beta_{n-1}}
\end{equation}

Where the product is the regular path attenuation accumulated from sampled albedos and $\beta_{n-1}$ is the product of sampling probabilities $Tr$ up until depth $n-1$ (see Algorithm~\ref{alg:termination}).

\subsection{Cache Training}
\begin{figure}[!htb]
    \centering
    \includegraphics[width=0.75\columnwidth]{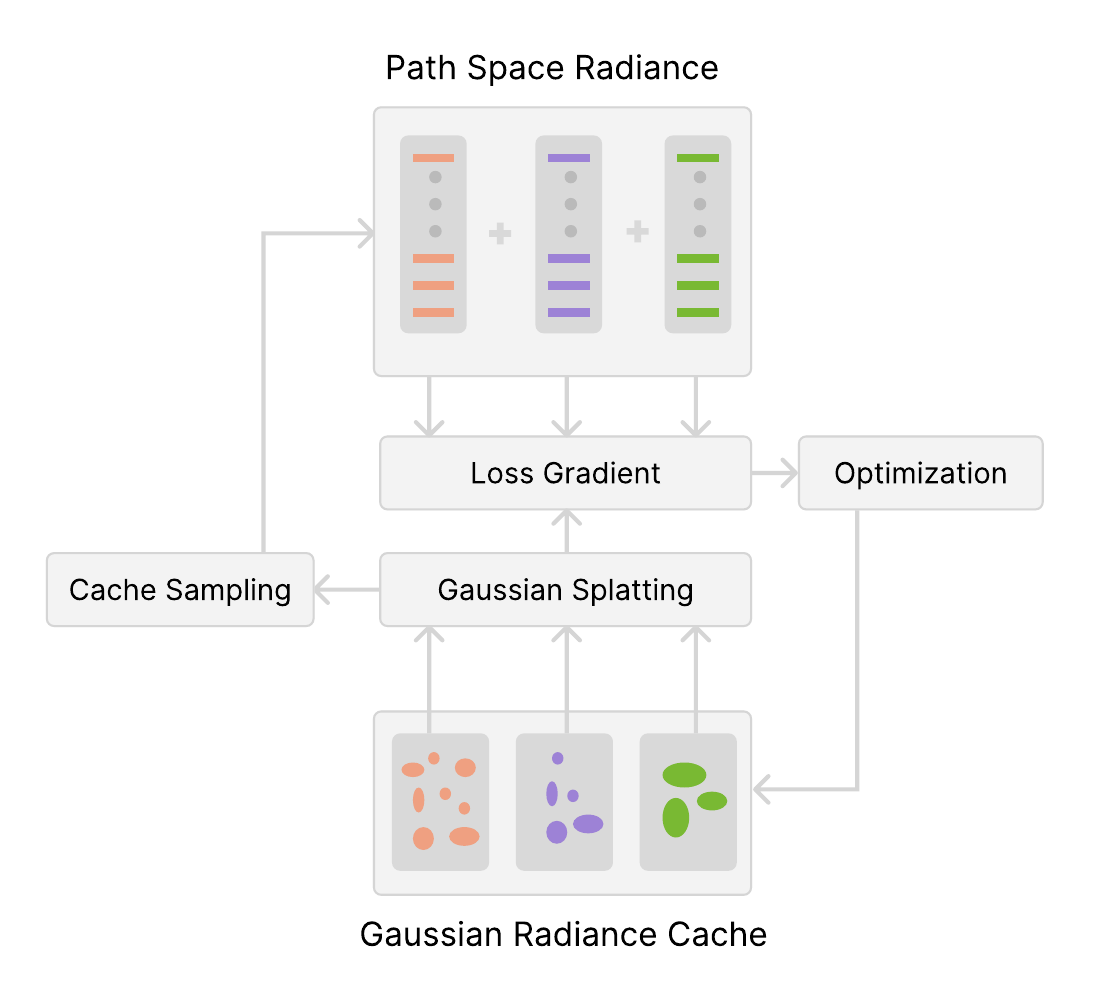}
    \vspace{-0.1in}
    \caption{Our cache is trained in real time on paths obtained during the training cycle. The optimizer uses Gaussian splatting and noisy path samples to optimize individual cache levels. At the same time, cached path radiance is sampled during rendering to improve image quality and runtime.}
    \label{fig:path_space_caching}
    \vspace{-0.1in}
\end{figure}

The radiance cache is trained in real time, and training can be toggled as needed. To train the cache, we collect unbiased samples from the renderer, attenuate them, and assign them to their appropriate cache level (Figure~\ref{fig:path_space_caching}). In the case of both natural and early path termination, the sample is simply attenuated by $Tr_{out}$ (see Algorithm~\ref{alg:termination}) and stored in an intermediate path buffer. We maintain $K$ intermediate path buffers---one for each level in the cache. At the time of termination, we determine the current path length, which determines the intermediate buffer that the sample gets assigned to.

At the end of one sample pass, each pixel will have at most one valid entry across the intermediate path buffers. At this time, the samples are attenuated, meaning that their total path throughput is taken care of. However, these samples are ultimately noisy as they represent the radiance from only one possible path of length $n$ through the volume. Conventionally, Gaussian splatting applications use clean target images, such as photographs, to calculate gradients for optimization. This poses a potential problem for our application.

We make the important observation that the expected value of samples in the buffers behaves analogously to regular path tracing, where paths of all lengths are collected indiscriminately. This means that if we were to take an infinite number of samples, we could accumulate a clean image for each of the buffers representing the radiance contribution in path subspace $n$.

With this observation, we apply the principle of learning clean data from noisy targets, which was first outlined by Lehtinen et al.~\cite{lehtinen18noise2noise}. This work shows that gradient descent optimization can generally be applied to problems with noisy targets if the corrupted target's expected value is the noise-free limit. Since this is the case for our path space buffers, we can apply this theory to enable training on noisy input buffers. This allows us to use the intermediate buffers as target images to compute the loss for the gradient calculation during the inverse splatting step of the Gaussian rasterization algorithm.

\subsection{Optimization and Hyperparameters}
The original implementation of 3D Gaussian splatting~\cite{kerbl3Dgaussians} includes various hyperparameters to steer the optimization to scenes with sparse input views. In our case, a lack of data is not the problem. Therefore, we can fine-tune the original algorithm to work more efficiently for our use case. In the following, we describe various additions and changes that we made to the optimization algorithm. 

\textbf{Spherical Harmonics.} The original 3D Gaussian splatting implementation~\cite{kerbl3Dgaussians} proposes the use of spherical harmonics 
(SH) to capture view-dependent effects such as specular reflections. While a reasonable addition for traditional computer vision tasks, Mallick et al.~\cite{mallick2024taming} show that a considerable amount of time is spent on the optimization of SH coefficients. Since view-dependent effects are not a driving factor in scientific volume visualization, we set the SH degree to $0$, which effectively reduces the optimization problem to three view-independent dimensions (RGB).

\revdel{
\textbf{Individual Learning Rates.} Since we generate a well-fit point cloud to initialize our cache levels, we rely less heavily on the algorithm's capability to optimize for Gaussian mean, covariance, and scale. In our experiments, we observed that more static Gaussians behave better when using noisy targets. We lower these individual learning rate settings to reduce their influence on the visual outcomes. 
}

\revadd{
\textbf{Densification and Pruning.} We do not use the original densification and pruning strategies introduced by Kerbl et al.~\cite{kerbl3Dgaussians}. The noise of our training samples makes the splitting, cloning, and pruning mechanisms too unstable for real-time training. Furthermore, removing these heuristic-based strategies improved not only stability but also the runtime of our method.
}

% \textbf{Optimizer.} We choose the AdamW~\cite{Loshchilov2017DecoupledWD} optimizer to fit the Gaussians. In our experiments, we noticed disrupting instabilities when using Adam. This becomes especially apparent if the training is enabled for a prolonged period of time while exposing training to a wide range of differing viewports. Using AdamW effectively regularizes the optimized parameters regardless of their gradient, preventing the emergence of extremes in scale, color, or orientation oscillation.

\textbf{Loss.} The data we are fitting has a high dynamic range (HDR) with potentially unbounded radiance values. This is in contrast to most scene representation tasks, which will draw samples from LDR sources like photographs. Commonly used loss terms like the L1 loss are ill-suited to adapt to HDR data as very bright samples generate disproportionately large gradients, which can overshadow the contribution of other samples. To address this, we adapt the HDR loss proposed by Lehtinen et al.~\cite{lehtinen18noise2noise}, which contains a normalization term and reads as follows.

\begin{equation}
    \mathcal{L}_{hdr} = \frac{(\hat{x} - \hat{y})^2}{k(\hat{y} + 0.01)^2}
\end{equation}
%1/k * ( (actual - expected) ^ 2 / (actual + 0.01) ^ 2 )

\noindent where $\hat{x}$ is the expected target value (i.e., the noisy path sample) and $\hat{y}$ is the predicted sample generated from the Gaussian splatting process. As inputs are images, we average the loss over all pixels $k$.

\textbf{Adaptive Learning Rates.} Each parameter type is assigned a separate learning rate. Please refer to our evaluation for specific starting values used in our trials. \revadd{Since we generate a well-fit point cloud to initialize our cache levels, we rely less heavily on the algorithm's capability to optimize mean, covariance, and scale.}
Furthermore, we employ a learning rate schedule that is reflective of the user's interaction with the rendering system. When there is no change in the viewport, we gradually reduce the learning rate to locally improve the quality of the cache. In practice, we scale the learning rate by the logarithm of the number of frames that the current viewport has been observed (Equation~\ref{eq:lr_schedule}).

\begin{equation}
    \eta_{t} = \eta_{0} \frac{1}{1 + \log(t)}
\label{eq:lr_schedule}
\end{equation}

If the viewport is changed, the learning rate is reset to its initial level, allowing for quick adaptation to changed inputs. The longer the viewer resides, the smaller the gradient-based updates become and the better the fit becomes locally, which improves the cache quality for the current viewport and its immediate neighbors. Although all learning rates are scaled in the same way, an adaptive algorithm that respects each parameter's gradients might be an interesting future addition.

\textbf{Regularization.} In contrast to most other scene representation approaches, our cache stores radiance values with unpredictably high variance. This impacts the performance of the optimization steps. Gradients can take on large values, which can occur very sparsely in high-variance scenarios. All this leads to instability in the training. In our tests, this specifically manifested as flickering and exploding Gaussians whose size quickly exceeded the scene scale and whose color dominated the image. To address this issue, we introduce regularization to penalize the emergence of large parameter values. Specifically, we use the AdamW~\cite{Loshchilov2017DecoupledWD} optimizer, which regularizes parameters independent of their adaptive step sizes. This modification ensures stability during training and prevents the aforementioned artifacts.

\subsection{Implementation}
We implement the radiance cache using C++ and CUDA. Optimization is handled via the PyTorch C++ API~\cite{PyTorch} and is directly integrated with the C++ runtime. We expose the caching library via a C-style API, which allows it to be easily adapted to various applications and languages. 
%Furthermore, we provide Python bindings for the API via pybind11~\cite{pybind11}.

For our evaluation, we integrate the cache into a scientific volume path tracer written in C++ using NVIDIA OptiX 7.3~\cite{OptiX} as a raytracing backend. Rendering kernels and caching integration are written in CUDA.
\section{Evaluation}

We evaluate our method in terms of runtime and image quality. 
Our results are compared against renderings using a baseline path tracer that implements uniform sampling and next-event estimation (NEE)\revadd{, all of which use an isotropic phase function}. \revadd{We also compare against state-of-the-art neural radiance caching (NRC)~\cite{muellernrc}.} Furthermore, we provide additional data and an ablation study in the supplemental material.

\begin{figure*}[!htb]
    \centering
    \begin{subfigure}[b]{.49\linewidth}
    \includegraphics[width=\linewidth]{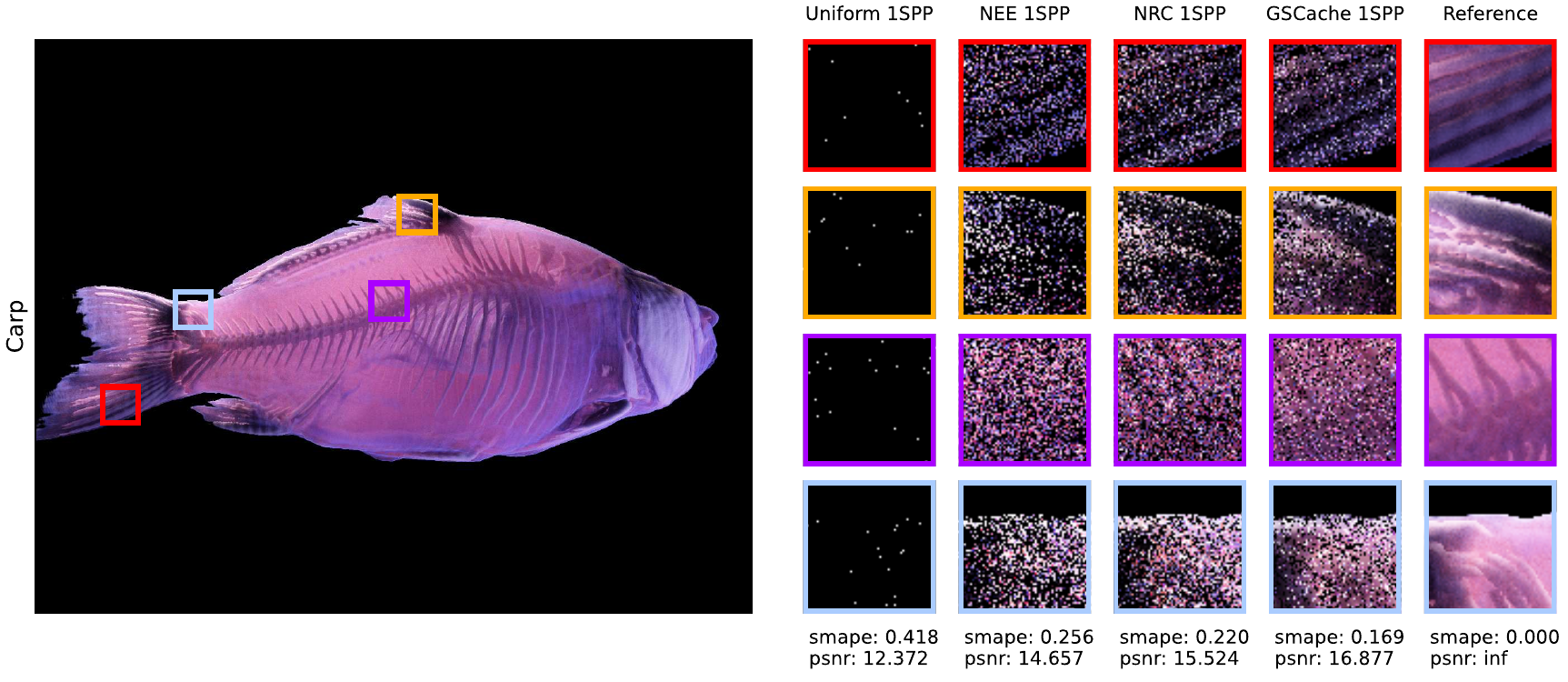}
    \end{subfigure}
    \begin{subfigure}[b]{.49\linewidth}
    \includegraphics[width=\linewidth]{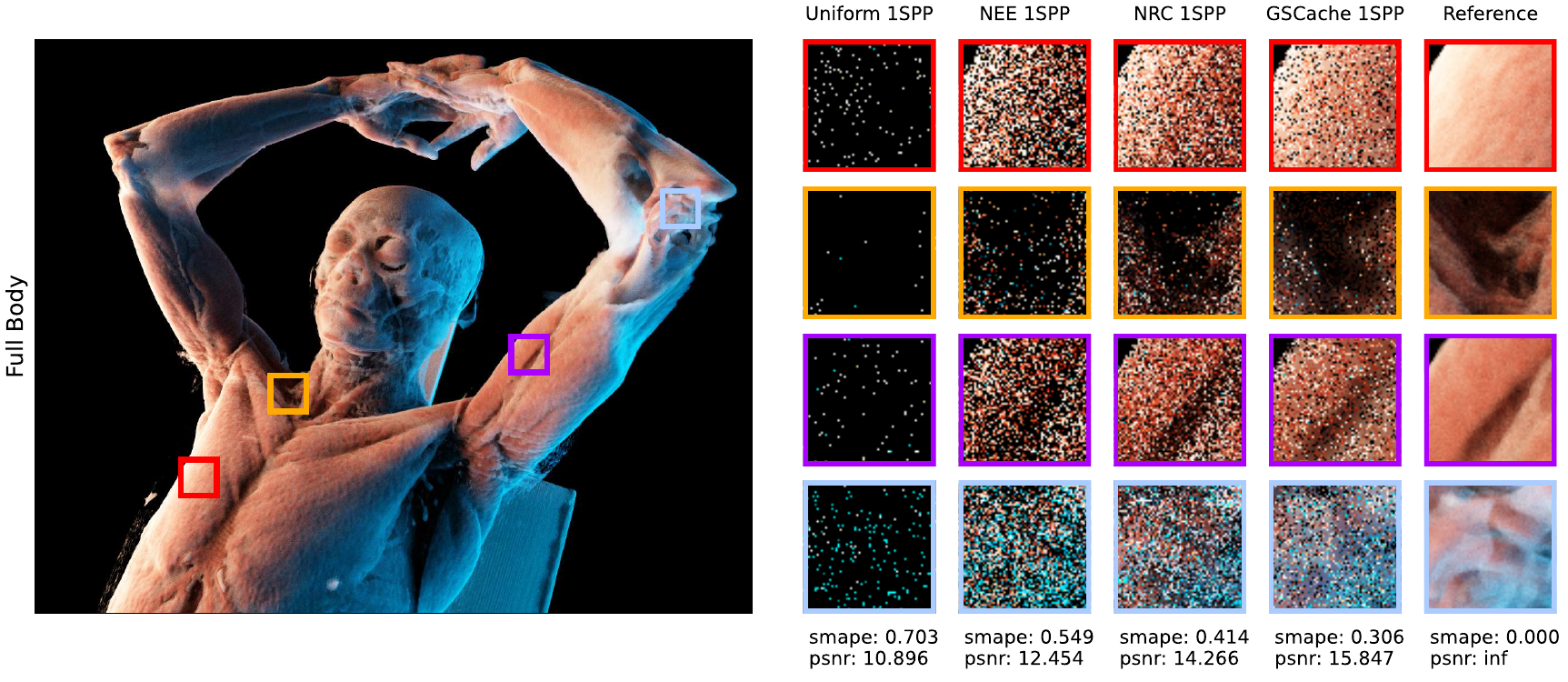}
    \end{subfigure}
    
    \begin{subfigure}[b]{.49\linewidth}
    \includegraphics[width=\linewidth]{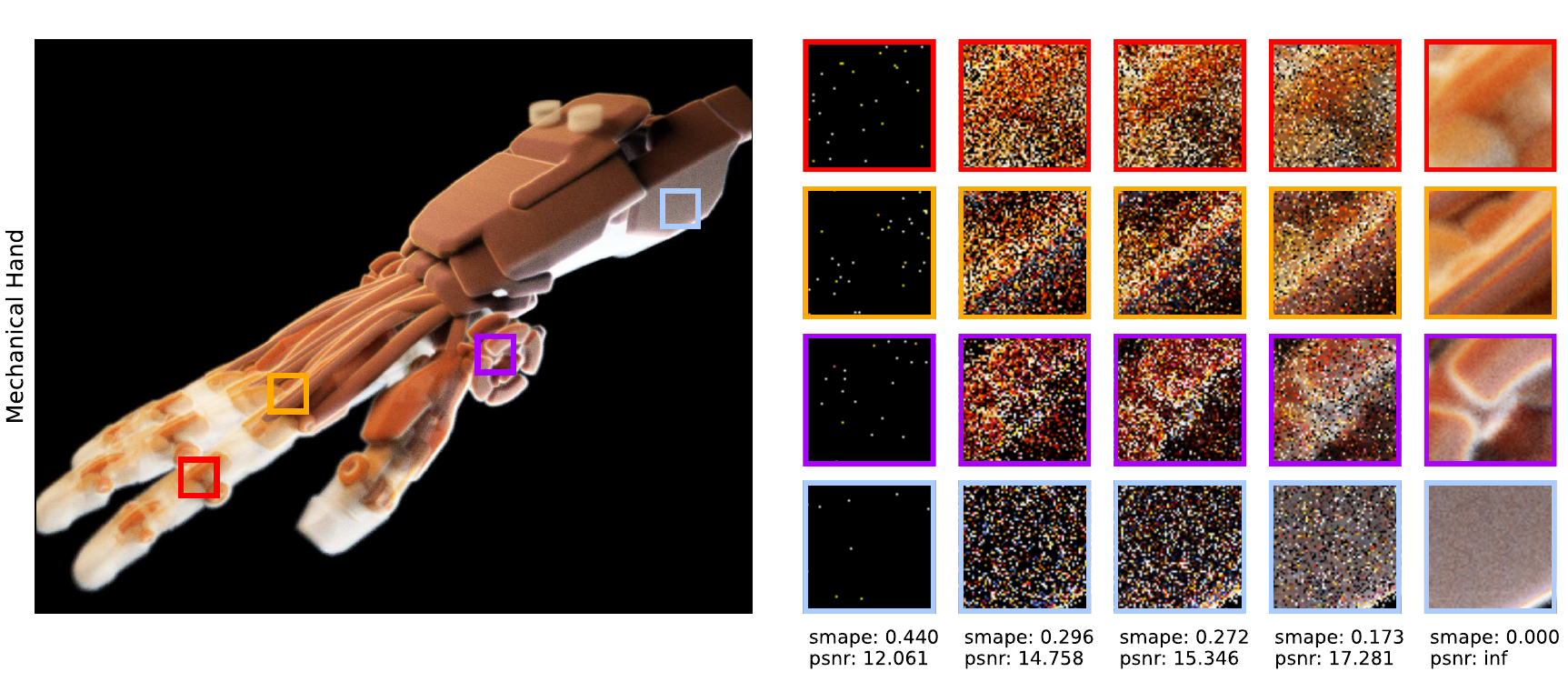}
    \end{subfigure}
    \begin{subfigure}[b]{.49\linewidth}
    \includegraphics[width=\linewidth]{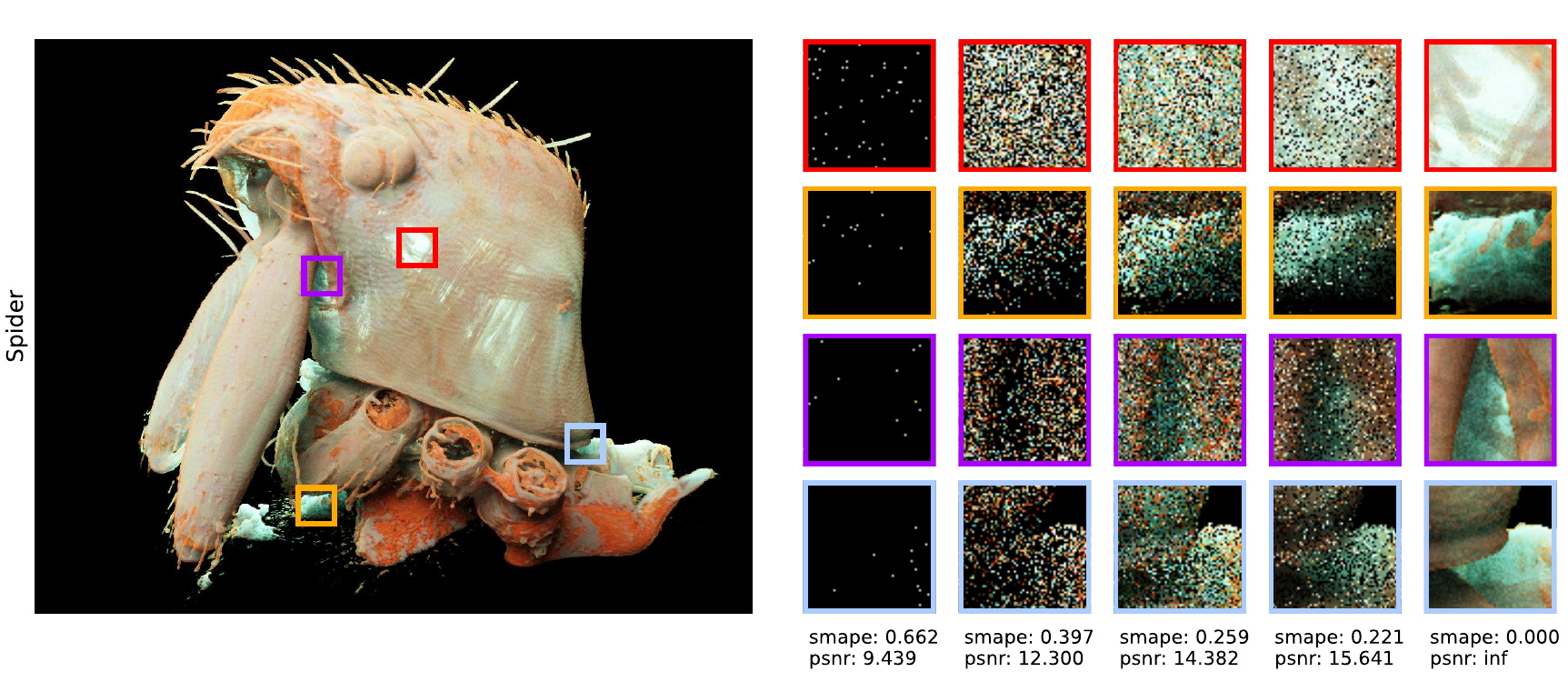}
    \end{subfigure}
    
    \begin{subfigure}[b]{.49\linewidth}
    \includegraphics[width=\linewidth]{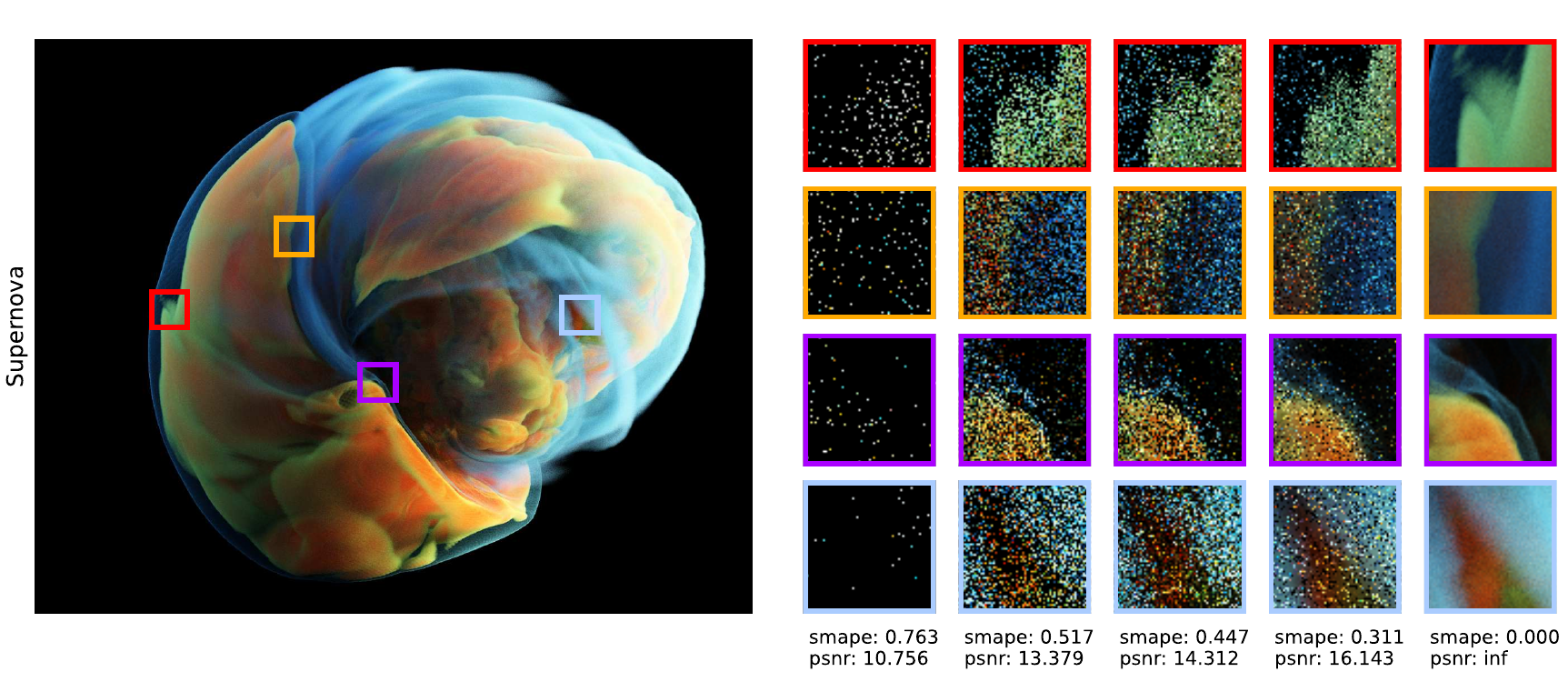}
    \end{subfigure}
    \begin{subfigure}[b]{.49\linewidth}
    \includegraphics[width=\linewidth]{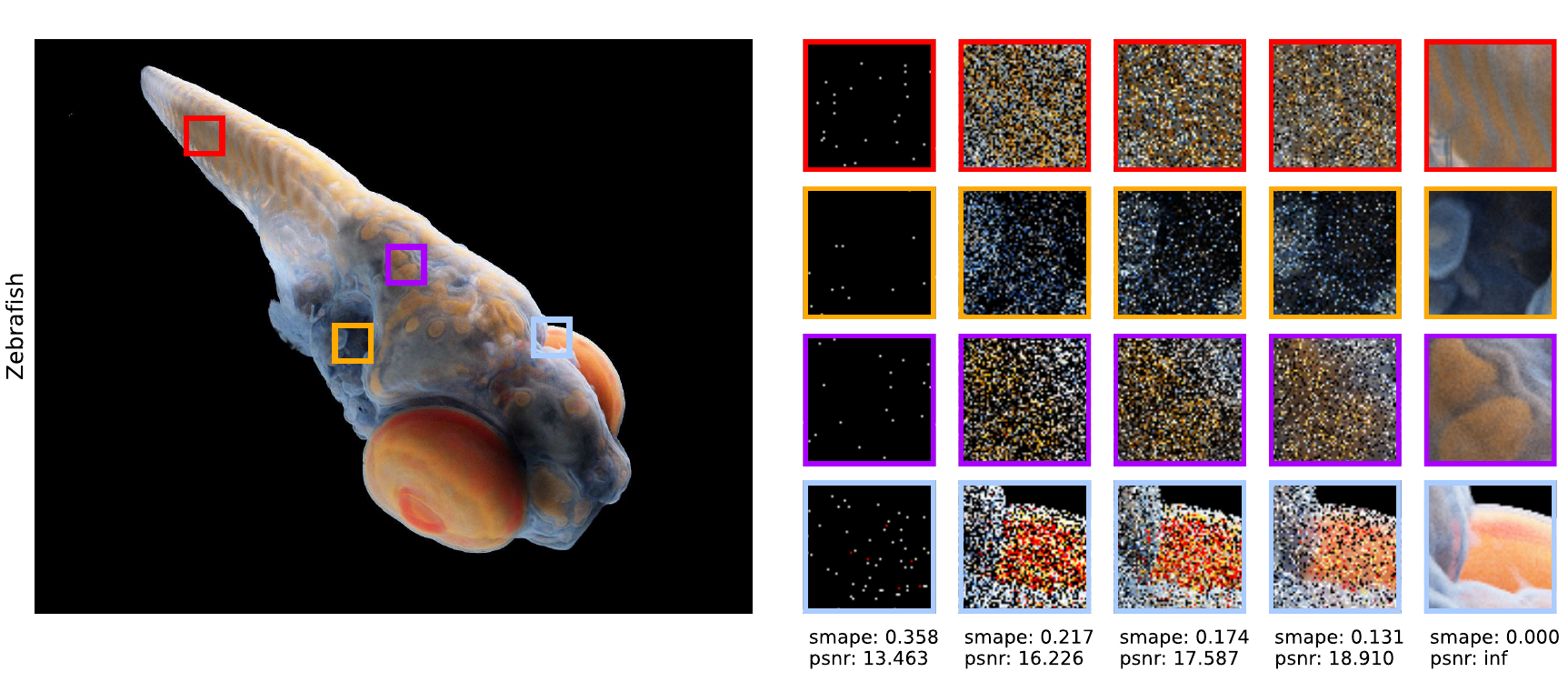}
    \end{subfigure}
    \caption{Visual quality of our method compared to the baseline path tracer. We show results for images at 1 SPP and compare our method (GSCache) against a baseline volume path tracer with uniform sampling (Uniform), a version that uses next-event estimation (NEE)\revadd{, and our implementation of NRC~\cite{muellernrc}}.}
    \label{fig:image_quality_matrix}
\end{figure*}

\subsection{System Specifications and Setup}
All tests were run on the same machine under equal conditions. The workstation used for these tests features an Intel Core i7-6900K CPU with 64GB of memory and an NVIDIA TITAN RTX with 24GB of video memory. The system was running Ubuntu 22.04.3 LTS, and all components were compiled using GCC 11.4.0 and NVCC 12.6.
In our evaluations, we use a common set of settings and parameters for our renderer and caching algorithm unless otherwise specified. Image resolution for performance evaluation was chosen at 720p and at 1280$\times$1024 for quality comparison to better frame datasets. Rendering was done at 1 sample per pixel (SPP) unless otherwise indicated.
We \revdel{generally} initialize the cache using $N=300$k \revdel{initial} samples with subsequent cache levels receiving an exponentially sub-sampled set of the initial points (see Section~\ref{sec:cache_init}). \revadd{In our setup, using three cache levels, this results in a total cache size of $29400000$ bytes (approx. $28$MB) for each scene and an average initialization time of $411.15$ms. See the supplemental material for a detailed analysis.}

For our tests, we use the following initial values for the learning rates (LR). Point position LR is $1.16e{-3}$, color LR is $1.25e{-2}$, rotation LR is $1e{-3}$, scaling LR is $0$, and opacity LR is $1.5e{-1}$. \revadd{We implement a version of NRC~\cite{muellernrc} for comparison to our method. To this end, we gather NEE samples as training data, use the same path termination heuristic as our method, and disable self-training for a fair comparison. Since volume radiance beyond the first bounce is highly diffuse, we use sample positions as input, as using spatio-directional inputs resulted in lower image quality in our tests. The model configuration is the same as NRC~\cite{muellernrc}. However, we use the hash-grid encoding for better positional encoding. The total size of the NRC cache in our setup is $284999680$ bytes (approx. $272$MB).}

\subsection{Datasets}
Our method is evaluated on a variety of volume datasets. In the following, we describe each dataset's characteristics (see Table~\ref{tab:datasets}).
We chose the datasets for this study to show a representative selection of data from medicine, biology, physics, and other fields. Since the present study concerns real-time applications, we did not consider extreme-scale datasets, which would require specialized data streaming and rendering techniques and might limit the system to offline rendering, weakening the justification for a cache. \revadd{For each of the datasets, we use a predefined 1D transfer function which remains unchanged throughout the evaluation.}

\begin{table}[!htb]
\caption{We use volume datasets from medical scans, CT scans, and scientific simulations to test our method. Data ranges in size and complexity.}
\centering
    \begin{tabularx}{\columnwidth}{l|X|X}
        \toprule
        \textbf{Dataset} & \textbf{Dimensions} & \textbf{Data Type} \\
        \midrule
        \textsc{FullBody} & $512\times512\times1299$ & uint8\\
        \textsc{MechanicalHand} & $640\times220\times229$ & float32\\
        \textsc{Supernova~\cite{ornl}} & $432\times432\times432$ & float32\\
        \textsc{Carp} & $128\times128\times256$ & uint8\\
        \textsc{Zebrafish} & $592\times413\times956$ & float32\\
        \textsc{Spider} & $957\times1195\times1003$ & uint16\\
        \toprule
    \end{tabularx}
\label{tab:datasets}
\end{table}

\subsection{Visual Quality}
We evaluate the visual quality of results obtained from our method by comparing them to uncached results. The baselines comprise results from a volume path tracer using uniform sampling and another version using next-event estimation. Each scene includes a single volume dataset with a custom transfer function (see Table~\ref{tab:datasets}). The scenes are lit by \revdel{several spherical area lights}\revadd{a two-point lighting setup using two spherical area light sources placed outside the volume. We choose small area lights to highlight the utility of radiance caching as scenes with ambient or environment lighting suffer less from sampling noise due to broader radiance coverage}.

\begin{figure}[!htb]
    \centering
    \includegraphics[width=0.70\columnwidth]{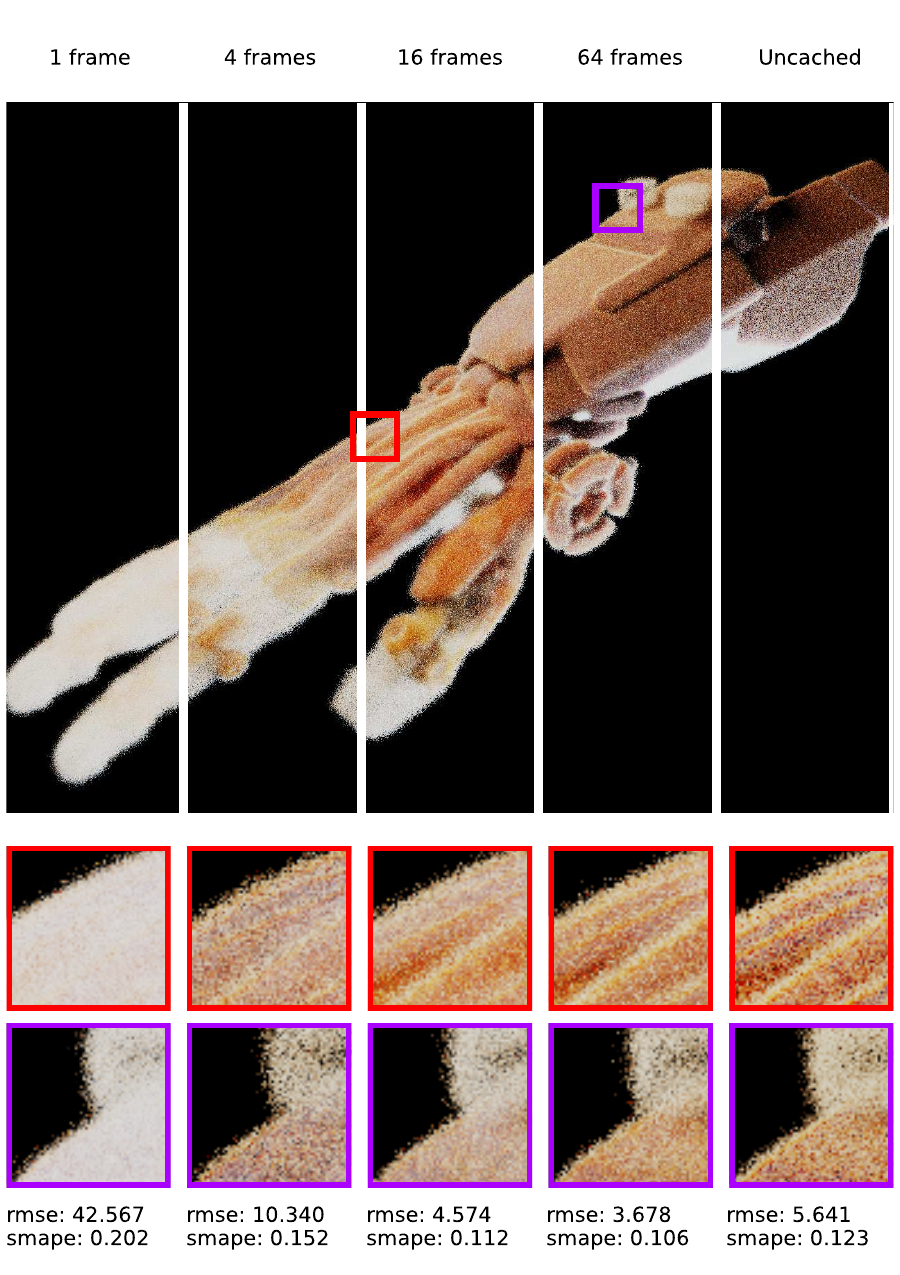}
    \caption{Comparison of cache state after a cold-start training for 64 frames. Frame 1 shows the freshly initialized state with raw albedo colors. Over the next few frames, the cache quickly adapts to the scene. We show the cache states after 1, 4, 16, and 64 frames and juxtapose them with an uncached frame capture. Images are rendered at 16 SPP, and quality metrics are computed compared to the unbiased reference image. Note that at 16 SPP, the cache provides diminishing returns as more unbiased paths are accumulated. Nevertheless, our cache outperforms the unbiased renderer at the same number of samples.}
    \label{fig:cache_training_progression}
\end{figure}

\begin{figure}[!htb]
    \centering
    \includegraphics[width=0.70\columnwidth]{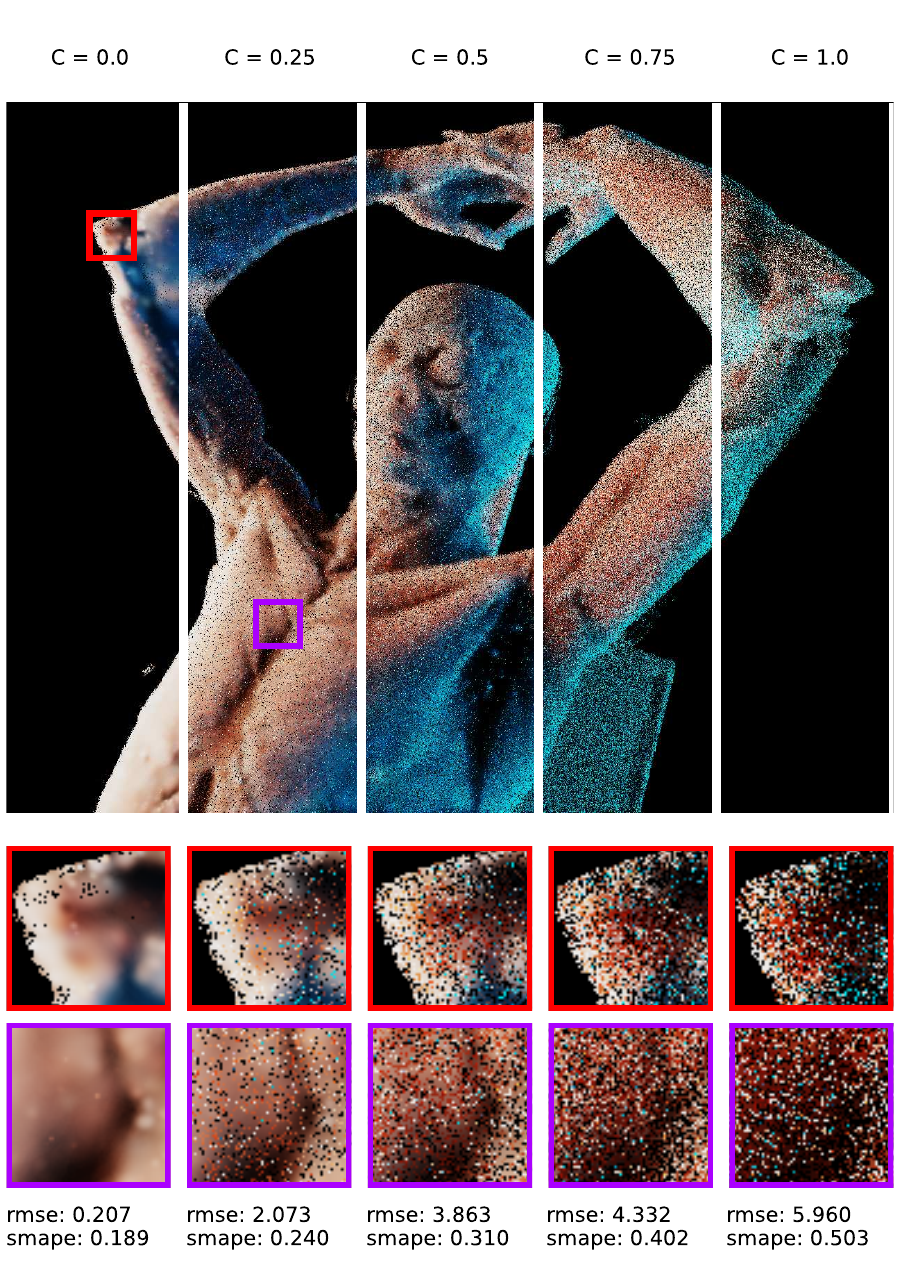}
    \caption{Comparison of using different values for the user-defined sampling coefficient $C$ at 1 SPP. Lower values favor termination into the cache, while higher values lead to longer paths. The figure shows the trade-off between bias and variance that can be made by choosing different values for $C$. While lower values for $C$ generally increase image metrics, they also introduce more bias into the final image.} 
    \label{fig:bias_variance_tradeoff}
\end{figure}

In Figure~\ref{fig:image_quality_matrix}, we compare the image quality of renderings from several different scenes. The results show that our method produces significantly higher image quality at low sample rates. Images are less noisy, as evidenced by the generally higher PSNR, and overall image quality \revdel{significantly}\revadd{notably} exceeds that of baseline \revadd{renderings and state-of-the-art caching~\cite{muellernrc} results}\revdel{renderings} at equivalent sampling rates.
We highlight the cache quality over time in Figure~\ref{fig:cache_training_progression}. After a cold start, it takes less than 16 samples for the cache to adapt to a point where it outperforms the baseline renderer in terms of visual quality. As rendering progresses, we significantly improve over the baseline and achieve superior image quality. Lastly, we investigate the effects of different cache sampling coefficients $C$, which we define in Section~\ref{sec:early_path_termination}. The results show how $C$ can be used to move along the bias-variance scale to find a suitable balance between image quality, rendering speed, and unbiasedness (Figure~\ref{fig:bias_variance_tradeoff}). In practice, the lower bound $C = 0.0$ is likely not going to be used as it corresponds to pure cache sampling without introducing unbiased paths, which implicitly precludes the possibility of cache training. We found a value of $C=0.5$ to be an adequate setting in our test scenes. Furthermore, we discuss the implications of different $C$ values for rendering performance in the next section.

\begin{table*}[!htb]
    \caption{Frame timings of our method on different datasets captured from a screen-filling camera fly-through of 200 frames. We allowed for a 40-frame warm-up period to initialize the cache and show path tracing time (PT) for all four methods and splatting time (ST) \revadd{, inference time (IT),} and optimization time (OT) for our method~\revadd{and NRC~\cite{muellernrc}}. \revadd{IT includes a composition pass to back-propagate and compose cached radiance.} The notation $PT_{nee*}$ denotes NEE with the addition of our path termination heuristic~\revadd{and $PT_{\text{nee}**}$ additionally adds path records and training sample collection for NRC~\cite{muellernrc}}. All values were captured at $C=0.5$ with $N=300$k initial points. All timings are in milliseconds.}
    \centering
    \begin{tabularx}{\textwidth}{l|X|X|X|X|X|X|X|X}
    \toprule
        \multirow{2}{*}{\textbf{{Dataset}}} & \multicolumn{2}{c|}{\textbf{Path Tracing}} & \multicolumn{3}{c|}{\textbf{Path Tracing + Ours}} & \multicolumn{3}{c}{\textbf{Path Tracing + NRC}}\\ 
        & {{$\text{PT}_{\text{uniform}}$}} &	 {{$\text{PT}_{\text{nee}}$ }} &	 {{$\text{PT}_{\text{nee}*}$}} & ST & OT & {$\text{PT}_{\text{nee}**}$} & IT & OT \\

    \midrule
    \midrule
    
% \textsc{FullBody} & 28.92 & 40.39 & 22.72 & 3.12 & 15.78\\
% \textsc{MechanicalHand} & 24.62 & 33.29 & 23.36 & 8.94 & 33.48\\
% \textsc{Supernova} & 23.21 & 42.80 & 28.85 & 10.29 & 13.38\\
% \textsc{Carp} & 41.17 & 65.47 & 37.86 & 14.31 & 23.78\\
% \textsc{Zebrafish} & 64.83 & 151.43 & 85.19 & 7.88 & 24.61\\
% \textsc{Spider} & 71.37 & 148.01 & 81.86 & 24.00 & 77.89\\
% \midrule
% \textbf{Overall} & \textbf{42.35} & \textbf{80.23} & \textbf{46.64} & \textbf{11.42} & \textbf{31.49}\\

\textsc{FullBody} & 28.92 & 40.39 & 22.72 & 3.12 & 15.78 & 32.00 & 4.29 & 29.74 \\
\textsc{MechanicalHand} & 24.62 & 33.29 & 23.36 & 8.94 & 33.48 & 32.51 & 3.95 & 26.85 \\
\textsc{Supernova} & 23.21 & 42.80 & 28.85 & 10.29 & 13.38 & 35.10 & 4.25 & 28.70 \\
\textsc{Carp} & 41.17 & 65.47 & 37.86 & 14.31 & 23.78 & 53.74 & 3.83 & 20.95 \\
\textsc{Zebrafish} & 64.83 & 151.43 & 85.19 & 7.88 & 24.61 & 120.19 & 3.12 & 22.10 \\
\textsc{Spider} & 71.37 & 148.01 & 81.86 & 24.00 & 77.89 & 119.61 & 4.18 & 26.31 \\
\midrule
\textbf{Overall} & \textbf{42.35} & \textbf{80.23} & \textbf{46.64} & \textbf{11.42} & \textbf{31.49} & \textbf{65.53} & \textbf{3.94} & \textbf{25.78}\\

    \toprule
    \end{tabularx}
    \label{tab:rendering_performance}
    \vspace{-0.2in}
\end{table*}

% \begin{figure}[!htb]
% \centering
% \begin{subfigure}[b]{.45\linewidth}
%     \includegraphics[width=0.95\columnwidth]{images/results/train0.pdf}
% \end{subfigure}
% \begin{subfigure}[b]{.45\linewidth}
%     \includegraphics[width=0.95\columnwidth]{images/results/bias0.pdf}
% \end{subfigure}
% \caption{Analysis of training and choice of sampling coefficient $C$. (Left) Comparison of cache state after a cold-start training for 64 frames. Frame 1 shows the freshly-initialized state with raw albedo colors. Over the next few frames, the cache quickly adapts to the scene. We show the cache states after 1, 4, 16, and 64 frames and juxtapose them with an uncached frame capture. Images are rendered at 16 SPP, and quality metrics are computed compared to the unbiased reference image. Note that at 16 SPP, the cache provides diminishing returns as more unbiased paths are accumulated. Nevertheless, our cache outperforms the unbiased renderer at the same number of samples. (Right) Comparison of using different values for the user-defined sampling coefficient $C$ at 1 SPP. Lower values favor termination into the cache, while higher values lead to longer paths. The figure shows the trade-off between bias and variance that can be made by choosing different values for $C$. While lower values for $C$ generally increase image metrics, they also introduce more bias into the final image.}
% \end{figure}

\begin{figure}[!htb]
    \centering
    \includegraphics[width=0.80\columnwidth]{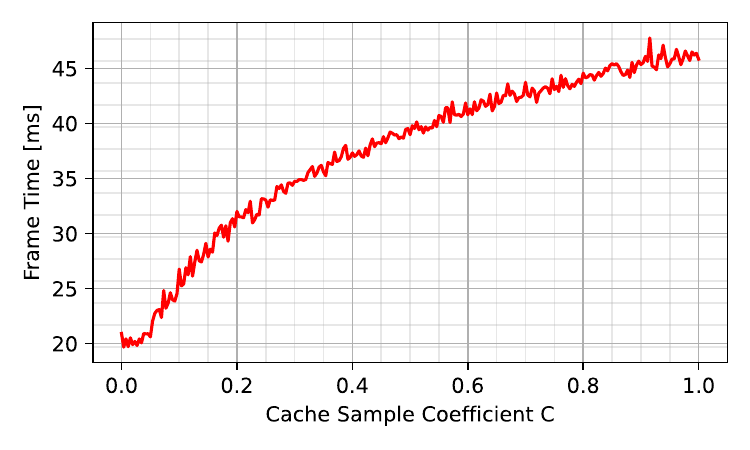}
    \vspace{-0.1in}
    \caption{Influence of the cache sampling probability $C$ on rendering times. Data was recorded on the \textsc{FullBody} dataset at 1 SPP.} 
    \label{fig:c_curve}
\end{figure}

\begin{figure}[!htb]
    \centering
    \includegraphics[width=0.80\columnwidth]{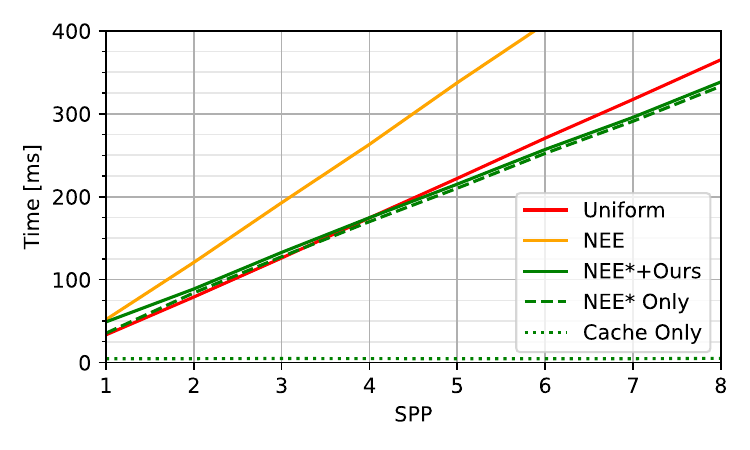}
      \vspace{-0.1in}
    \caption{Cache amortization behavior example from the \textsc{FullBody} dataset. NEE* denotes NEE times with our ray termination heuristic. At low sample counts, our method performs better than the baseline NEE path tracer. As SPP increase, the time to generate the cache buffers stays constant, and the overall runtime for our method grows more slowly than both baseline approaches.} 
    \label{fig:runtime_spp}
\end{figure}

% \begin{figure}[!htb]
%     \centering
%     \begin{subfigure}[b]{.45\linewidth}
%     \includegraphics[width=\linewidth]{images/results/highlight0.pdf}
%     \end{subfigure}
%     \begin{subfigure}[b]{.45\linewidth}
%     \includegraphics[width=\linewidth]{images/results/highlight0.pdf}
%     \end{subfigure}
%     \caption{\todo{Write caption when figure is final}}
%     \label{fig:image_quality_highlight}
% \end{figure}

\subsection{Runtime Performance}
Runtime performance is an important factor for the use of a radiance cache, as we generally want to achieve higher image quality in the same time or less compared to the uncached application.

We provide itemized runtime numbers in Table~\ref{tab:rendering_performance}. Results show that~\revadd{timings are overall comparable to NRC~\cite{muellernrc} and that} while there is a runtime overhead to rendering the cache every frame, we generally make up for that time by cutting longer paths short and terminating them into the cache. Naturally, the sampling constant $C$ influences the exact performance gain from shorter paths (see Figure~\ref{fig:c_curve}). We found a value of $C=0.5$ to be a good balance between image quality, bias, and runtime performance (see Figure~\ref{fig:bias_variance_tradeoff}). \revadd{Note that the number of lights and choice of transfer function can impact rendering performance overall but does not have any meaningful impact on caching times as we use the same number of cache points regardless of the lighting environment and do not employ splitting. This means the user-defined cache size stays constant throughout the application life cycle.} Training times can make up for a significant portion of the splatting-related workload, which can be alleviated by selectively enabling training when there are significant changes in the scene or the viewport. For example, training could be disabled if the learning rate crosses a lower limit threshold due to a user residing in the same viewport for prolonged periods of time or the optimization rate could be adjusted (i.e., varying the number of frames after which the optimization routine is invoked) based on the differential change in camera position and viewing angle to avoid optimizing for viewports that might be outdated in just a few frames. We note that our implementation is not highly optimized, leaving potential performance improvements for the optimization step by leveraging custom low-level implementations. Please refer to our discussion for potential performance improvements suggested by Mallick et al.~\cite{mallick2024taming} to improve optimization and rasterization speeds.

One interesting aspect of the cache is that\revadd{, unlike NRC~\cite{muellernrc},} its runtime overhead is constant in the number of samples taken per pixel. So, while increasing SPP generally diminishes the returns of a cache as more unbiased samples are collected, we can still benefit from its use. Each sample can access the same cache state to improve sample quality and shorten paths. This results in an overall performance gain as each sample is generated faster due to early termination, while the cache state only needs to be rendered once per frame, regardless of SPP. We show a runtime curve from one of our datasets in Figure~\ref{fig:runtime_spp}, which outlines how the cache usage amortizes over the number of samples.

% \subsection{Ablation}
% \todo{Only write this if there is space.}
% To show the effectiveness of our design we perform ablation studies over the cache depth, size, and hyperparameters. In the following, we present results for each of these categories.
\section{Limitations and Future Directions}
In the following, we discuss the limitations of our method and highlight areas for potential future work.

\textbf{Multi-Pass Rendering and Runtime Optimizations.}
Our implementation of the cache uses different sets of Gaussians for each cache level. These levels are entirely independent of each other. This simplifies the implementation and usage of the cache but leads to several inefficiencies in the rasterization and optimization pipeline, as each of these stages needs to be executed separately for each cache level. A possible optimization and extension of this work could be to develop a joint rasterization and optimization pipeline that can jointly render and optimize multiple sets of Gaussians while still producing their independent contributions and gradients for use in the caching application. Furthermore, at the time of writing, differentiable 3D Gaussian splatting is still a relatively new topic, with ongoing research into improving storage demands as well as splatting and optimization times. One example of such developments is a recent work by Mallick et al.~\cite{mallick2024taming}, who present multiple times faster optimization speeds at similar visual quality. As more research is published on these topics, we anticipate the value of our cache design to further increase.

\textbf{Cache Size.}
During rendering and optimization, the cache resides entirely on the GPU. One downside of Gaussian-splatting-based approaches is the relatively high memory requirement compared to other scene representation techniques. This can negatively impact application performance as the caching pipeline might take up a non-negligible fraction of GPU memory during runtime. There have been several recent attempts to cull the memory footprint of Gaussian splatting applications~\cite{mallick2024taming,lee2024compact3dgs,niedermayr2024compressed3dgs,papantonakis2024memfoot3dgs}. Extensions of our approach could include such cache compression and reduction methods to create a smaller and more efficient cache. 
% As a side effect, such an optimization will likely result in some performance gain during runtime, as fewer Gaussians need to be optimized and rasterized.

\textbf{Adaptivity Versus Consistency.}
For a radiance cache to be effective, it needs to react quickly to changes in the scene by evicting or forgetting past states and adapting to new ones. \revdel{On the other hand,}\revadd{At the same time,} if conditions are static, we want the cache to provide consistent data to ensure image quality and temporal stability between frames. In this work, we develop an adaptive learning rate scheduler that is tied to the renderer and user interaction patterns to determine the optimal learning rate at any given point. For our purposes, this constitutes an adequate trade-off between the adaptability and consistency of our cache. However, this method is specifically tailored to our scientific visualization use case and might not be adequate for other applications. In such cases, there is a need to develop more sophisticated methods to establish a balance between adaptability and consistency in the cache. \revadd{Furthermore, while our design is largely agnostic to transfer function changes, it requires re-initialization for cases that alter the overall morphology of the visible volume. Future work could address refitting strategies for such cases.}
\revadd{Another consequence of the cache's high adaptivity is its short-term memory. If a viewport is not covered for some time, the Gaussians encoding the radiance for this viewport might be re-used to encode more recent views, thus ``forgetting'' views that have not been active for some time and preventing a global convergence of the cache in the long term. While we did not observe unrecoverable cases of view-dependent overfitting, this limitation requires continuous training while scene or viewport changes are expected. Future work could address this issue by focusing optimizations on Gaussians that are closest and most relevant to the current viewport, potentially using spatial subdivision and nearest-neighbor methods to find optimization candidates to facilitate longer-term convergence of the cache.}

\textbf{Bias.}
As with most radiance caching approaches, using our cache results in biased rendering results. \revdel{On the one hand, w}\revadd{W}e achieve very high image quality this way\revdel{. On the other hand,}\revadd{, however} there are no guaranteed error bounds for the error associated with bias. This is a known limitation of caching approaches. In our experiments, we observed tolerable amounts of bias-related errors, which speaks to our cache's reconstruction quality. Nevertheless, if unbiased results are a requirement, our method is not suitable. In those cases, users might fare better employing advanced sampling methods like ReSTIR~\cite{restir} or Volume ReSTIR~\cite{volumerestir}.

\revdel{
\subsection{Applicability for Denoising}
One aspect of our cache is that each cache level approximates the expected value of the path sub-space of length $n$. Many modern image denoisers for path tracing now increasingly rely on kernel prediction to filter noisy images. Applying a kernel to a pixel calculates a weighted sum of its neighborhood. Kernel weights are not static but are generally predicted by a neural network. Since our cache represents the expected value, its output can be interpreted as a perfectly sampled path. Since such a path perfectly lies within the distribution of possible paths, its value is suitable for use in kernel-based denoisers. This fact makes our cache a viable means to produce higher-quality inputs to real-time image denoisers, further improving the utility of real-time path tracing for scientific visualization.

\subsection{Surface Geometry}
Currently, our method is intended specifically for improving volume rendering applications. However, the method's design is general enough to be extended to a broad range of applications, including surface and implicit geometry. The main challenge is finding appropriate initialization techniques for these new data types, which involves computing a set of seed points to avoid cold-starting the cache from random points and providing meaningful initial conditions. With such additions, our cache should work out of the box for any type of scene configuration.
}

\textbf{Self-Training and Cache Design.}
Currently, our cache is trained in path space purely on unbiased path samples obtained during rendering. This limits training data to fully concluded paths. Future work could explore the potential of reusing cached entries for paths $\hat{n} < n$ by demodulating them with the throughput $\sigma_i$ and sampling probability $Tr$ of prior segments to generate samples for earlier cache levels in the spirit of M\"uller et al.~\cite{muellernrc}. Furthermore, recent advances in ray tracing Gaussian representations~\cite{condor2025dontsplat,moenne20243dgrt,wu20253dgut} allow for arbitrary ray queries which can enable world-space cache designs similar to NRC~\cite{muellernrc}. %This modification would improve the effective number of available training samples per frame allowing faster and potentially more stable convergence.% which in turn might help incorporate further optimizations to improve splatting performance or image quality.

\section{Conclusion}
We introduce a novel method for radiance caching using path-space radiance samples in combination with a multi-level hierarchy of Gaussians. The properties of Monte Carlo noise allow us to learn sub-space path radiance from noisy inputs using gradient descent. 
Our evaluation results show that GSCache helps volume path tracers converge faster and produce higher-quality results in a shorter amount of time. At the same time, our framework is easy to integrate into existing applications due to its reliance on path-space radiance, unlike other popular caching methods, which usually require extensive information about the sampling process, intersection metadata, and more. Caching radiance in path-space also has the advantage of being agnostic to the characteristics of the scene. It can handle surfaces just as easily as volumetric data and other representations like signed distance fields or implicit neural representations. With this work, we hope to encourage further research into the use of Gaussian splatting for scientific visualization applications.

%% if specified like this the section will be omitted in review mode
%% TODO: Add back in in final version
% \acknowledgments{This research is sponsored in part by the National Institute of Health under grant no. P41 EB032840, National Science Foundation under grant no. IIS-2427770, and the Intel oneAPI Center of Excellence.}
\acknowledgments{This research is supported in part by the U.S. National Science Foundation with grant III-2427770 and the Intel oneAPI Center of Excellence.}

\bibliographystyle{abbrv-doi-hyperref}

\bibliography{template}

\appendix % You can use the `hideappendix` class option to skip everything after \appendix
%% Supplemental Sections 
\section{Ablation Study}
We provide an ablation study to showcase the effect of various choices regarding the initialization and training of our cache. In the following, we investigate how initialization size, spacing, and scaling of Gaussians, and various training hyperparameters influence the final quality of the images.

\subsection{Cache Size}
\begin{figure*}[h]
    \centering
    \includegraphics[width=1\linewidth]{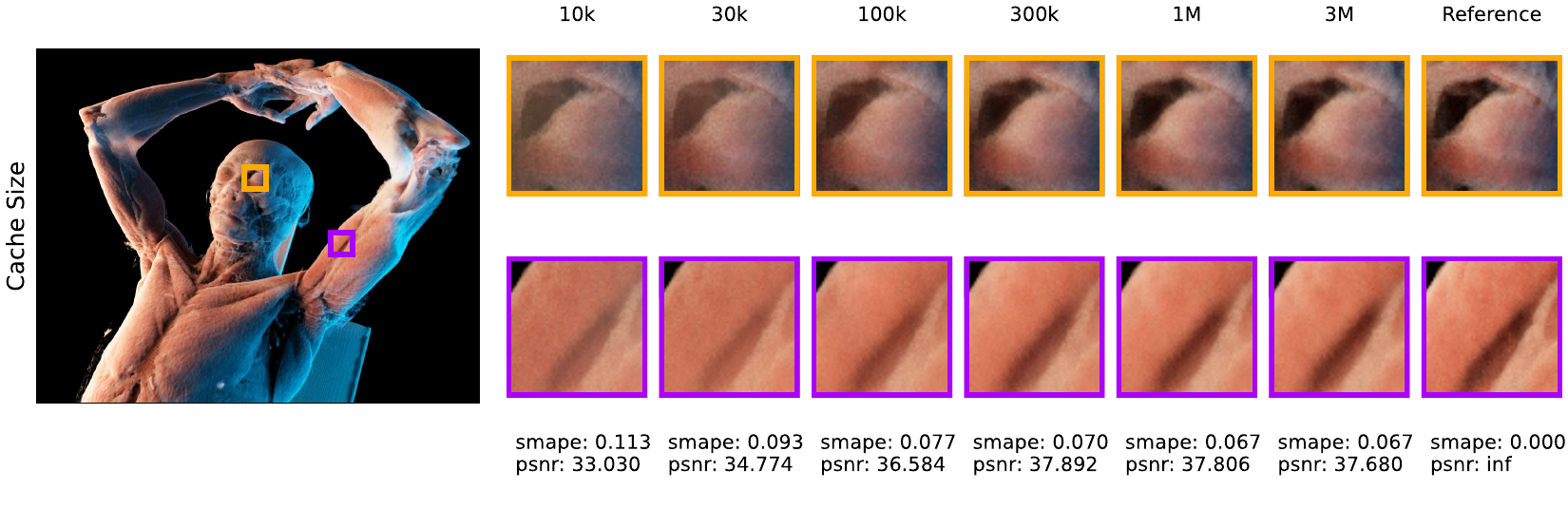}
    \caption{Comparison of the visual quality of our method when using different cache initialization sizes $N$ as shown on the \textsc{FullBody}. The columns show images captured from different runs at $1024$ SPP using an initial cache size noted at the top of each column. Results show significant gains in quality up to around $N=300$k, after which additional gains become negligible.}
    \label{fig:initsize}
\end{figure*}
In this study, we compare the image quality of our results when using different cache initialization sizes $N$. The results in Figure~\ref{fig:initsize} indicate that from $N = 10$k to about $N = 300$k we see significant improvements in image quality. Beyond this size, the additional gains are marginal considering the increased processing and storage cost of higher numbers of Gaussians. This informed our choice of $N=300$k in our other experiments. Note that $N$ denotes the number of Gaussians in the first cache level and that subsequent levels follow the schedule we describe in the main document, in which each level is half the size of its predecessor. The lower cache sizes generally produce more washed-out looks in the final image. This is because fewer Gaussians are forced to represent larger areas of the image, which reduces local quality and adaptability to finer details. At $N=3$M, this effect becomes almost unnoticeable. We note that the exact cache size threshold needed to produce acceptable results can vary from dataset to dataset simply due to their size. The \textsc{FullBody} dataset shown in Figure~\ref{fig:initsize} was one of the larger datasets in our tests, which is why we chose it for this ablation.

\subsection{Hyperparameters}
\begin{figure*}[h]
    \centering
    \includegraphics[width=1\linewidth]{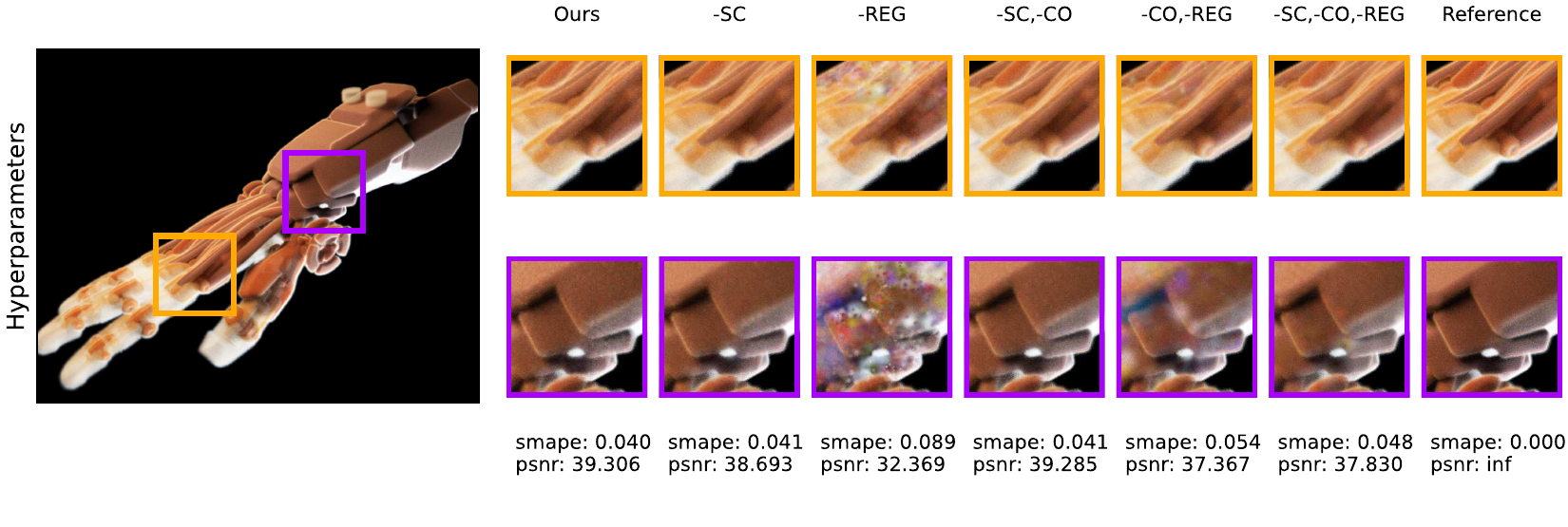}
    \caption{Ablation of our hyperparameter choice. The image compares results from our configuration with various modified versions of those settings. Data was captured by first rotating the camera around the object for 256 frames, followed by a recuperation period of another 256 frames in a still viewport. We present different combinations of settings with different choices removed from our final settings. The labels in the figure denote the following modifications. (-SC) removes the initial size cap on Gaussians. (-CO) removes our limitation on the scaling learning rate and allows scaling at a LR of $0.0125$. Lastly, (-REG) removes the regularization during training.}
    \label{fig:hyperparams}
\end{figure*}
To understand how our hyperparameter choice influences final image quality, we conduct an ablation over several key components in our pipeline. Figure~\ref{fig:hyperparams} summarizes our findings. In this study, we test different combinations of hyperparameter settings, gradually removing constraints or replacing key components with alternatives. Specifically, we consider the removal of the size gap and downscaling of the initial set of Gaussians that we describe in the main document (Figure~\ref{fig:hyperparams} (SC)). Furthermore, we allow optimization for Gaussian covariance (CO), which by default we disabled for training stability. Lastly, we test how the regularization influenced training stability by replacing AdamW~\cite{Loshchilov2017DecoupledWD} with a regular Adam~\cite{kingma2015adam} optimizer. 

The results indicate that regularization has the highest impact on final cache quality. Without it, the cache fails to adapt to new viewports and gradients quickly explode, leaving us with numerous artifacts that make the cache practically unusable (see Figure~\ref{fig:hyperparams} Ours vs. -REG). This is in contrast to Kerbl et al.'s~\cite{kerbl3Dgaussians} original implementation, which worked well without regularization. Allowing the optimization to scale Gaussians by setting the scaling LR to a non-zero value can help offset training instability in cases where the optimizer fails to do so (see Figure~\ref{fig:hyperparams} -REG vs. -CO,-REG). We found this choice to make little difference when using regularization, and it is up to the specific application to determine the best setting. Finally, the initial size cap and downscaling of Gaussians helped preserve some of the finer details in the structure of the dataset while trading this advantage for a slightly noisier look (see Figure~\ref{fig:hyperparams} Ours vs. -SC). Overall, the image quality with size cap and downscaling proved to be slightly better. In summary, we see that our choices worked well for our set of test data, but we acknowledge that their impact on the final image quality and training performance might differ depending on the specific application, and we encourage users to experiment with different settings. Aside from that, a valuable insight we derive is that without regularization, real-time optimization on noisy data does not seem to be a viable option.
\revadd{
\section{Memory Footprint}
We briefly mention the memory footprint of our method in the main document. Here, we provide a more detailed analysis in Table~\ref{tab:memory_footprint}. The cache components refer to the different parameters stored for Gaussian splats. All data is stored as 32-bit floating-point values, which leaves further room for optimization. All sizes are reported in megabytes.

\begin{table}[!htb]
\caption{Breakdown of the memory footprint of our cache per component. Here $N$ is the size of each cache level, which is $300$K, $150$K, and $75$K, respectively, for our setup.}
\centering
    \begin{tabularx}{\columnwidth}{l|l|X|X|X}
        \toprule
        \textbf{Comp.} & \textbf{Dim.} & \textbf{L0 Size} & \textbf{L1 Size} & \textbf{L2 Size} \\
        \midrule
        \textsc{Position} & $N\times3$ & $3.43$ MB & $1.72$ MB & $0.86$ MB  \\
        \textsc{Rotation} & $N\times4$ & $4.58$ MB & $2.29$ MB & $1.14$ MB  \\
        \textsc{Color} & $N\times3$ & $3.43$ MB & $1.72$ MB & $0.86$ MB  \\
        \textsc{Scale} & $N\times3$ & $3.43$ MB & $1.72$ MB & $0.86$ MB  \\
        \textsc{Opacity} & $N\times1$ & $1.14$ MB & $0.57$ MB & $0.28$ MB  \\
        \midrule
        \textsc{Total} & $N\times14$ & $16.02$ MB & $8.01$ MB & $4.01$ MB  \\
        \toprule
    \end{tabularx}
\label{tab:memory_footprint}
\end{table}
}
\revadd{
\section{Cache Initialization Time}
The cache is initialized at the beginning of the application by randomly sampling the volume to generate a point cloud that follows the density distribution of the volume and the chosen transfer function. To this end, we use the same sampling method~\cite{woodcock1965} as in the main path tracer. In Table~\ref{tab:initialization_times}, we report average initialization timings per dataset. As our analysis shows, this process takes too much time to be repeated every frame, but it is not so slow as to prohibit re-initialization of the cache upon transfer function changes.

\begin{table*}[t]
\caption{Measurement of average initialization timings over five runs per dataset or a total of $30$ initializations with $N=300$K. The individual timing depends on the dataset size and average volume density as mapped by the transfer function.}
\centering
    \begin{tabularx}{\textwidth}{l|X|X}
        \toprule
        \textbf{Dataset} & \textbf{Initialization Times [ms]} & \textbf{Average Time [ms]}\\
        \midrule
        \textsc{Carp} &$315.688$, $328.575$, $325.539$, $309.039$, $317.080$ & $319.18$ \\
        % \midrule
        \textsc{FullBody} & $415.402$, $410.654$, $416.495$, $442.563$, $453.676$ & $427.76$ \\
        % \midrule
        \textsc{MechanicalHand} & $296.847$, $279.348$, $305.291$, $286.633$, $290.239$ & $291.67$\\
        % \midrule
        \textsc{Spider} & $314.366$, $316.603$, $310.582$, $321.163$, $307.771$ & $314.10$\\
        % \midrule
        \textsc{Supernova} & $428.015$, $440.562$, $438.888$, $432.853$, $414.349$ & $430.93$\\
        % \midrule
        \textsc{Zebrafish} & $684.607$, $678.135$, $675.923$, $685.786$, $691.880$ & $683.27$\\
        \midrule
        \textsc{Total} & - & $411.15$\\
        \toprule
    \end{tabularx}
\label{tab:initialization_times}
\end{table*}

}
\section{Image Quality}
\begin{figure*}[t]
    \centering
    \includegraphics[width=\linewidth]{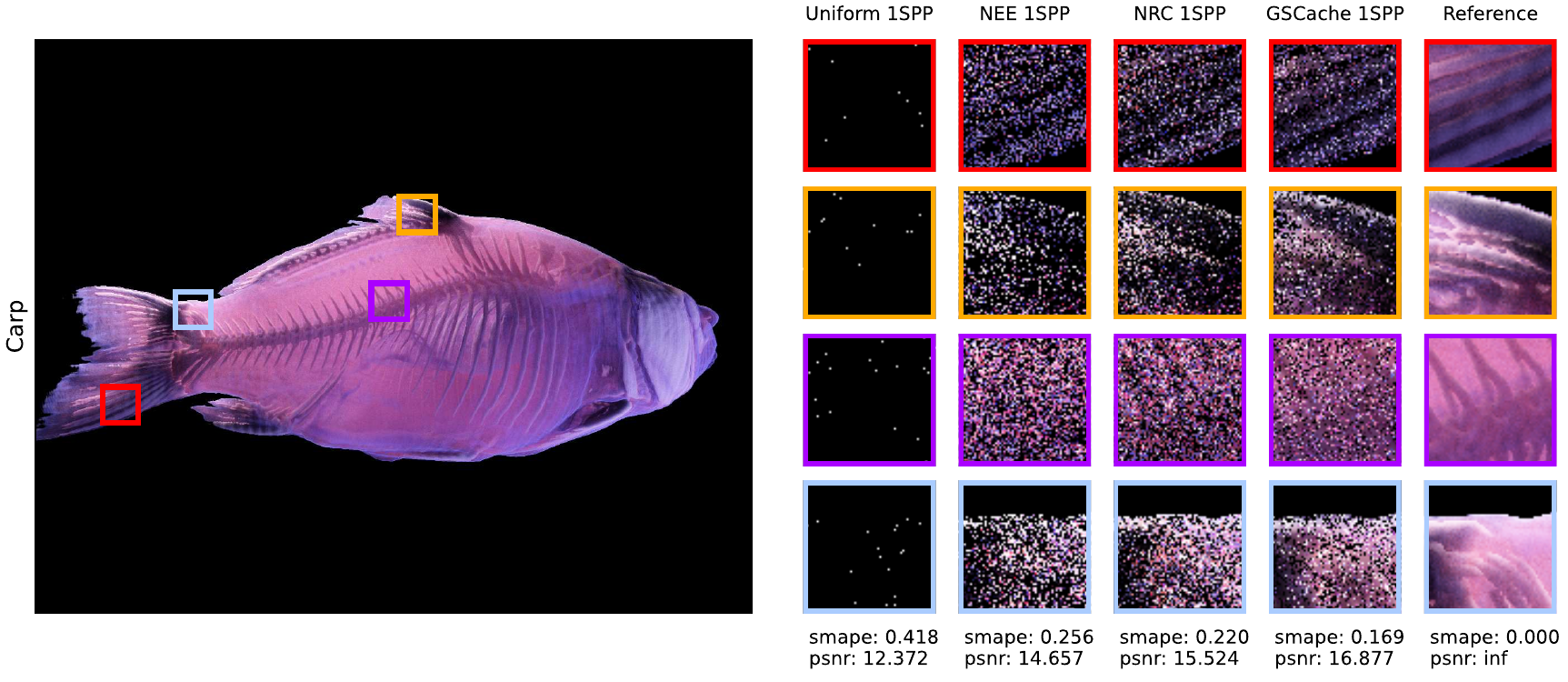}
    \caption{Visual quality of our method on the \textsc{Carp} dataset compared to the baseline path tracer. We show results for images at 1 SPP and compare our method (GSCache) against a baseline volume path tracer with uniform sampling (Uniform) and a version that uses next-event estimation (NEE).}
    \label{fig:image_quality_0}
\end{figure*}

\begin{figure*}[t]
    \centering
    \includegraphics[width=\linewidth]{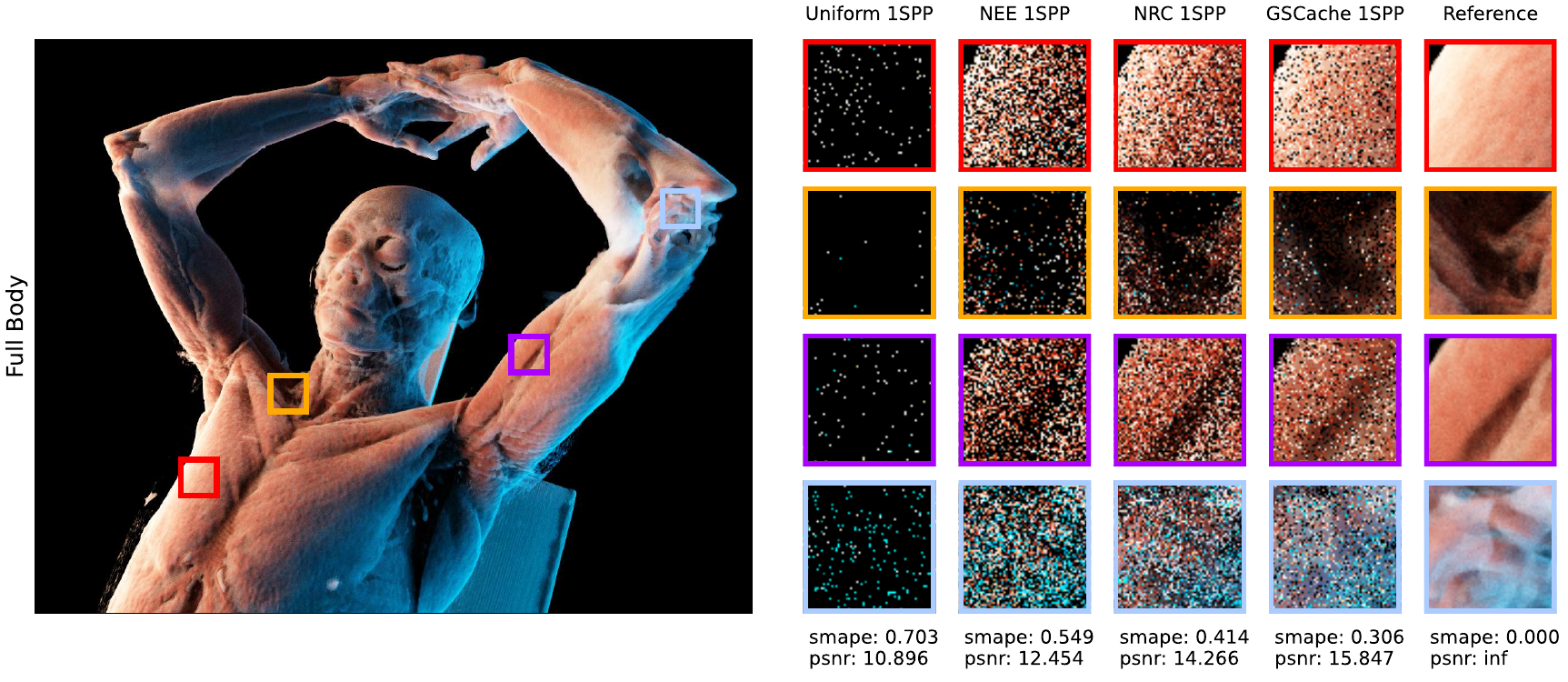}
    \caption{Visual quality of our method on the \textsc{FullBody} dataset compared to the baseline path tracer. We show results for images at 1 SPP and compare our method (GSCache) against a baseline volume path tracer with uniform sampling (Uniform) and a version that uses next-event estimation (NEE).}
    \label{fig:image_quality_1}
\end{figure*}

\begin{figure*}[t]
    \centering
    \includegraphics[width=\linewidth]{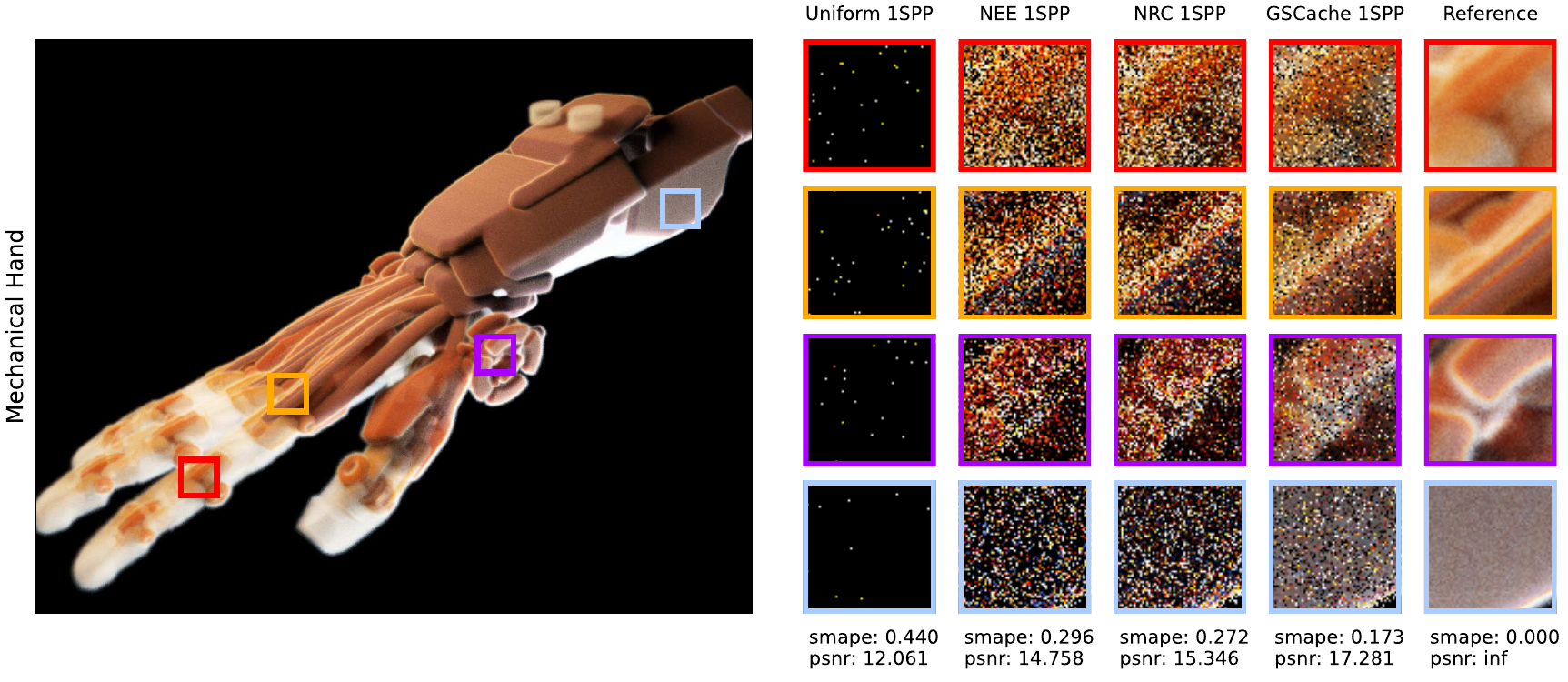}
    \caption{Visual quality of our method on the \textsc{MechanicalHand} dataset compared to the baseline path tracer. We show results for images at 2 SPP and compare our method (GSCache) against a baseline volume path tracer with uniform sampling (Uniform) and a version that uses next-event estimation (NEE).}
    \label{fig:image_quality_2}
\end{figure*}

\begin{figure*}[t]
    \centering
    \includegraphics[width=\linewidth]{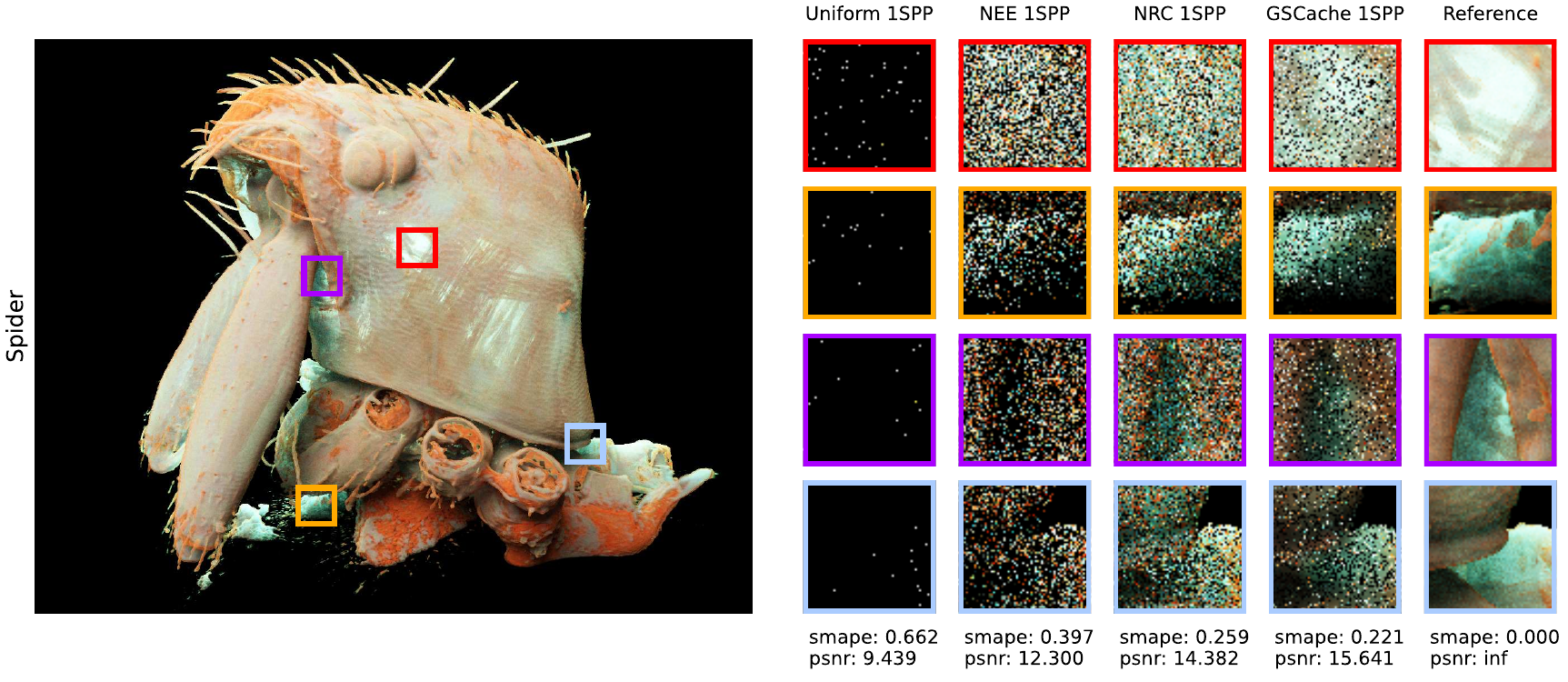}
    \caption{Visual quality of our method on the \textsc{Spider} dataset compared to the baseline path tracer. We show results for images at 1 SPP and compare our method (GSCache) against a baseline volume path tracer with uniform sampling (Uniform) and a version that uses next-event estimation (NEE).}
    \label{fig:image_quality_3}
\end{figure*}

\begin{figure*}[t]
    \centering
    \includegraphics[width=\linewidth]{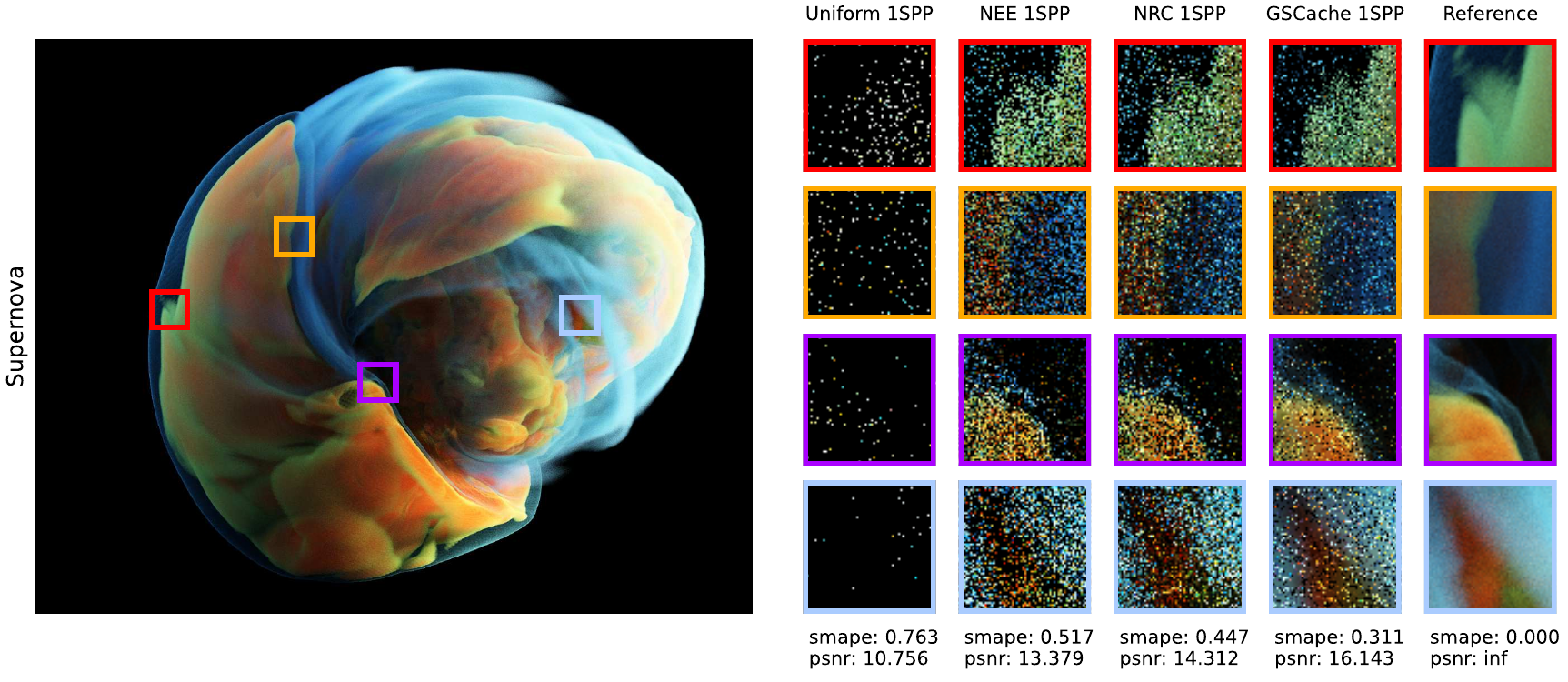}
    \caption{Visual quality of our method on the \textsc{Supernova} dataset compared to the baseline path tracer. We show results for images at 1 SPP and compare our method (GSCache) against a baseline volume path tracer with uniform sampling (Uniform) and a version that uses next-event estimation (NEE).}
    \label{fig:image_quality_4}
\end{figure*}

\begin{figure*}[t]
    \centering
    \includegraphics[width=\linewidth]{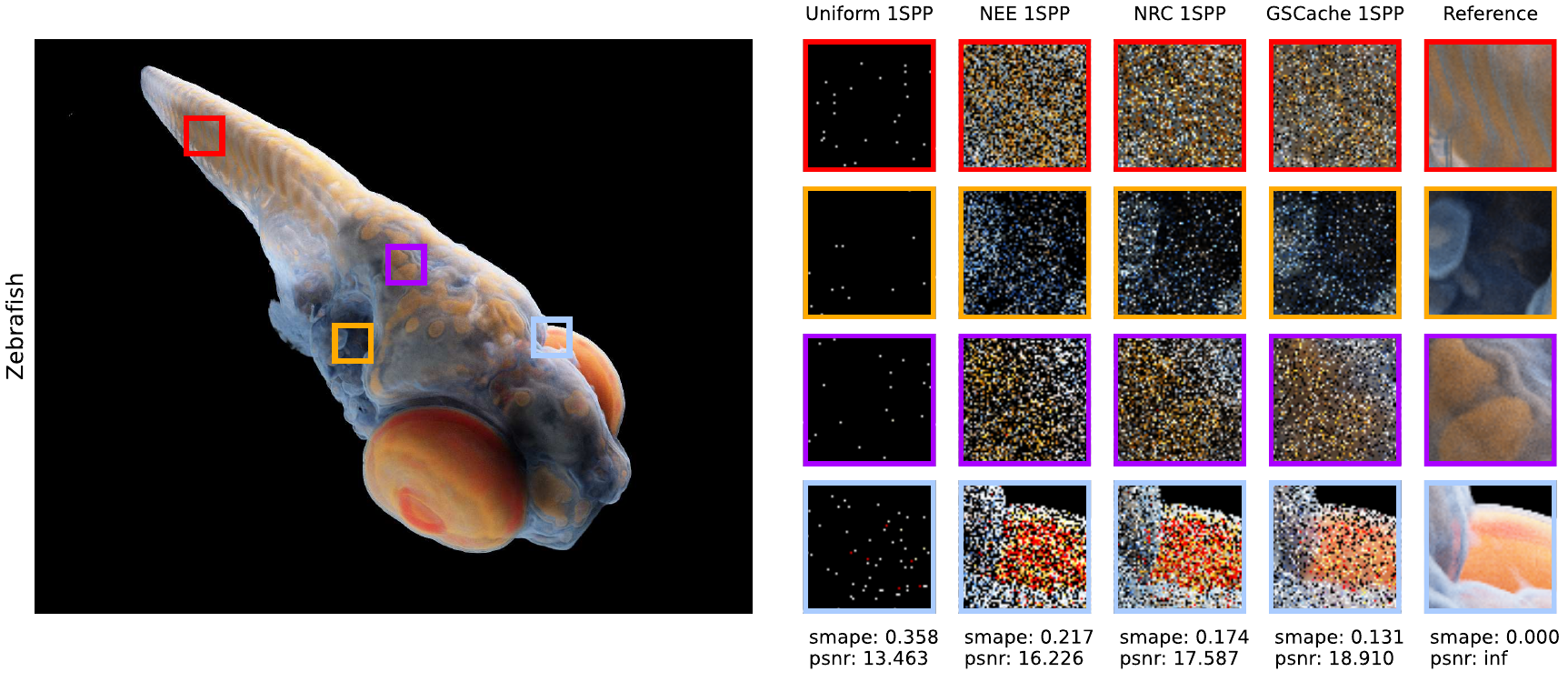}
    \caption{Visual quality of our method on the \textsc{ZebraFish} dataset compared to the baseline path tracer. We show results for images at 1 SPP and compare our method (GSCache) against a baseline volume path tracer with uniform sampling (Uniform) and a version that uses next-event estimation (NEE).}
    \label{fig:image_quality_5}
\end{figure*}

To complement the figures in the main document, we provide larger versions of the qualitative results in Figures~\ref{fig:image_quality_0},~\ref{fig:image_quality_1},~\ref{fig:image_quality_2},~\ref{fig:image_quality_3},~\ref{fig:image_quality_4}, and~\ref{fig:image_quality_5}.


\end{document}